\documentclass{aa}  
\usepackage{natbib}
\usepackage{graphics,graphicx}
\usepackage{txfonts}
\usepackage{ulem}
\usepackage{threeparttable}
\usepackage{hyperref}
\usepackage{multirow}
\usepackage{longtable}
\usepackage{supertabular}
\usepackage[switch]{lineno} 
\usepackage{csquotes}
\usepackage{placeins}

\makeatletter
\renewcommand*\aa@pageof{, page \thepage{} of \pageref*{LastPage}}
\makeatother

\begin{document} 

\def\gammajtff{J1045.3$+$2751}
\def\gammajtfn{J1049.8$+$2741}
\def\gammarxs{J1725.5$+$1312}
\def\gammasfe{J0658.6$+$0636}
\def\gammasfs{J0656$+$0539}
\def\whsp{1WHSP\,J104516.2$+$275133}
\def\txsA{TXS\,0506$+$056}
\def\pksa{PKS\,B1424$-$418}
\def\txsB{TXS\,1100$+$122}
\def\nvssA{NVSS\,J104938$+$274212}
\def\icnBone{WISEA\,J111439.67$+$122503.7}
\def\icnBtwo{WISEA\,J105553.74$+$103446.5}
\def\rxs{1RXS\,J172314.4$+$142103}
\def\pks{PKS\,1725$+$123}
\def\nvssB{NVSS\,J065844$+$063711}
\def\wisea{WISEA\,J065633.43$+$053922.7}
\def\nvssC{NVSS\,J065916$+$055252}
\def\icntxs{IC\,170922A}
\def\icnA{IC\,190704A}  
\def\icnB{IC\,200109A}  
\def\icnC{IC\,201021A}  
\def\icnD{IC\,201114A}  
\def\pflux{\mathrm{ph}\,\mathrm{cm}^{-2}\,\mathrm{s}^{-1}}
\def\Xflux{\mathrm{erg}\,\mathrm{cm}^{-2}\,\mathrm{s}^{-1}}
\def\deg{$^\circ$}
\def\fermi{{\it{Fermi}}}

   \title{Observing the inner parsec-scale region of candidate neutrino-emitting blazars}

   \author{C. Nanci  
        \inst{1,2},
   M. Giroletti 
        \inst{2},
   M. Orienti 
        \inst{2},
   G. Migliori
        \inst{2},
    J.  Mold\'on 
      \inst{3},
    S. Garrappa 
    \inst{4},
    M. Kadler
    \inst{5},
    E.   Ros     
       \inst{6},
       S.  Buson
     \inst{5},
    T.  An
        \inst{7},
   M. A. P\'erez-Torres
     \inst{3},
    F. D'Ammando 
      \inst{2},
    P. Mohan
      \inst{7},
    I.   Agudo
    \inst{3},
    B. W.   Sohn
    \inst{8},
    A.J. Castro-Tirado \inst{3},
     Y. Zhang  \inst{7}
      }
        
   \institute{Dipartimento di Fisica e Astronomia, Universit\`a di Bologna, via Gobetti 93/2, 40129 Bologna, Italy\\
               \email{cristina.nanci@inaf.it}  
        \and
             Istituto Nazionale di Astrofisica,   Istituto di Radioastronomia (IRA), via Gobetti 101, 40129 Bologna, Italy
        \and
           Instituto de Astrof\'isica de Andaluc\'ia (CSIC), Glorieta de las Astronom\'ia, s/n, E-18008 Granada, Spain 
        \and 
         Deutsches Elektronen-Synchrotron, 15738 Zeuthen, Germany
          \and
        Institut f\"{u}r Theoretische Physik und Astrophysik, Universit\"{a}t W\"{u}rzburg,
Emil-Fischer-Str. 31, 97074 W\"{u}rzburg, Germany 
          \and
        Max-Planck-Institut f\"ur Radioastronomie, Auf dem H\"ugel 69, 53121 Bonn, Germany
        \
        \and 
         Key Laboratory of Radio Astronomy, Shanghai Astronomical Observatory, Chinese Academy of Sciences, 80 Nandan Road, Shanghai 200030, China
         \and
        Korea Astronomy and Space Science Institute, 776 Daedeok-daero, Yuseong-gu, Daejeon 34055, Korea
       }
   \date{}
 
  \abstract
  {
 Many questions on the nature of astrophysical counterparts of high-energy neutrinos remain unanswered. There is increasing evidence of a connection between blazar jets and neutrino events, with the flare of the $\gamma$-ray blazar \txsA{} in  spatial and temporal proximity of \icntxs{} representing one of the most outstanding associations of high-energy neutrinos with astrophysical sources reported so far.
  }
  {With the purpose of characterising potential blazar counterparts to high-energy neutrinos,    we   analysed the parsec-scale regions of $\gamma$-ray blazars in spatial coincidence with high-energy neutrinos,  detected by the IceCube  Observatory. Specifically, we intended to investigate peculiar radio properties of the candidate counterparts  related to the neutrino production, such as  radio flares coincident to the neutrino detection or features in  jets morphology (limb brightening, transverse structures).} 
  {We  collected multi-frequency  very-long-baseline interferometry (VLBI) follow-up observations of candidate counterparts of four high-energy neutrino events detected by  IceCube  between January 2019 and November 2020, with a focus    on   $\gamma$-ray associated objects.  
  We analysed their radio characteristics  soon after the neutrino arrival in comparison with  archival VLBI observations and low-frequency radio observations. We discussed our results with respect to previous statistical works  and studies  on  the case of \txsA{}. 
    }
  {We identified and analysed in detail five potential neutrino emitting blazars. Our results suggest an enhanced state of activity for one source, \pks{}.
 However,  the lack of  adequate monitoring prior to the neutrino events was a limitation   in tracing  radio activity and morphological changes in all the  sources.}
   {We suggest that \pks{} is a promising  neutrino source candidate. For the other sources, our results alone do not  reveal a strong connection between the radio activity state at the neutrino arrival. 
   A  larger number of VLBI and multi-wavelength follow-up observations of neutrino events  is now  essential to understand the   neutrino production mechanisms in astrophysical sources. 
  }
   

 \titlerunning{Observing the inner parsec-scale region of candidate neutrino-emitting blazars}

 \authorrunning{C. Nanci et al.}

\maketitle

\section{Introduction}

The detection of the first neutrinos with PeV energy \citep{2013Sci...342E...1I,2013PhRvL.111b1103A} has revived the long-standing debate about the possible production site of these high-energy particles. Because energetic neutrinos are isotropically distributed across the sky \citep{2014PhRvL.113j1101A}, extragalactic sources are the first natural electromagnetic counterparts to look at.
Emission of $\gamma$-rays is also expected in neutrino production chains. However, the sub-degree/degree angular resolution of current $\gamma$-ray and neutrino detectors usually prevents unambiguous identification of the neutrino $\gamma$-ray counterpart. Since the extragalactic GeV-TeV population is dominated by a class of active galactic nuclei (AGN), the blazars \citep{2015ApJ...810...14A}, these have been extensively studied as the main contributors to the astrophysical neutrino flux. So far, we know from stacking analysis that the population of $\gamma$-ray blazars should contribute no more than 27\% of the diffuse neutrino flux  \citep{2017ApJ...835...45A}.

Blazars are radio loud AGN with a relativistic jet aligned with the observer's line of sight \citep{1995PASP..107..803U}.
Their strong outflow radiates over the entire electromagnetic spectrum, from the radio band to $\gamma$-rays. The emission from blazar jets is enhanced and blue-shifted due to relativistic effects and is characterised by a two-hump spectral energy distribution (SED).
The interaction between relativistic electrons and the magnetic field within the jet produces synchrotron emission at low frequencies, with a peak between $\sim10^{12.5}$ Hz and $\sim10^{18.5}$ Hz \citep{2012A&A...541A.160G}.
The peak of the high-energy emission is observed between $\sim$10$^{20}$ Hz and
$\sim$10$^{26}$ Hz \citep{2012A&A...541A.160G}. Both Inverse Compton (IC) and hadronic processes are capable of describing this second bump, but hybrid lepto-hadronic mechanisms have also been proposed \citep[e.g.][]{2019Galax...7...20B}.
In particular, in hadronic processes also neutrinos are produced.

Based on their optical properties, blazars are divided into flat spectrum radio quasars (FSRQs), which have strong emission lines from clouds around the AGN accretion disc (called broad line region, BLR); and BL Lacs (BLLs), which have weak emission lines or completely featureless optical spectra. According to the position of the synchrotron emission peak, $\nu_{\mathrm{peak}}$, blazars are also divided into  Low Synchrotron Peaked (LSP, or Low peaked BLL, LBL, relative to BLLs, $\nu_{\mathrm{peak}} < 10^{14}$ Hz), Intermediate Synchrotron Peaked (ISP/IBL, $10^{14}$ Hz $< \nu_\mathrm{peak} < 10^{15}$ Hz) and High Synchrotron Peaked (HSP/HBL, $\nu_{\mathrm{peak}} > 10^{15}$ Hz) \citep{2010ApJ...716...30A}.

Relativistic jets of  blazars represent the ideal particle accelerators in which high-energy neutrinos could come out as products of highly accelerated protons interacting with the environment \citep[e.g.,][]{1995APh.....3..295M}. 
This theoretical expectation was successfully confirmed in 2017, when it was identified the first (and so far only) significant evidence for a blazar-neutrino association. The IceCube-170922A (\icntxs{}) event, detected with an energy of 290 TeV and a high probability of being of astrophysical origin, was found in spatial coincidence with the blazar \txsA{} and in temporal coincidence with a $\gamma$-ray flare from this source, detected by the \textit{Fermi} gamma-ray space telescope and the MAGIC telescopes \citep{IceCube2018a}. The multi-wavelength campaign, which began after the discovery of the neutrino and the $\gamma$-ray flare, showed that \txsA{} emission increased in all bands of the electromagnetic spectrum, including the radio band.


A neutrino event  was previously tentatively associated with the bright $\gamma$-ray flaring blazar \pksa{} \citep{Kadler2016}. Unfortunately, due to the large angular uncertainty in the neutrino position ($\sim$15\deg $-$20\deg), the association of \pksa{} with the neutrino event is statistically less robust than in the case of \icntxs{} \citep[detected within a 90\% error region of less than 1\deg,][]{IceCube2018a}. Recently, \citet{2020ApJ...893..162F} and \citet{2021ApJ...912...54R} identified a promising connection between the FSRQ PKS\,1502+106 and the IC\,190730A event. This source was also found in a high state in the radio band and in a low state at $\gamma$-rays  \citep{2021ApJ...911L..18K}.
The possible explanations on the neutrino production arising from these  few observational hints are still incomplete and require further investigations of the possible neutrino counterparts.
 
Very long baseline interferometric (VLBI) observations  played a crucial role in modelling the neutrino production mechanism in \txsA{}. The milliarcsecond (mas) angular resolution achievable with VLBI provides a unique tool to penetrate deep into the parsec scale of the jets. This is the site where particle acceleration is expected and where radio flares typically originate.
Using VLBI archival and post-event data of \txsA{}, \citet{2020A&A...633L...1R} showed that a rapid expansion of the radio core occurred after the neutrino detection. They also interpreted the limb brightening in the jet of \txsA{} as a sign of transverse velocity structure.  
This reinforced the idea of a connection with the neutrinos production given the considerations from previous theoretical studies  \citep[e.g.,][]{2014ApJ...793L..18T}. \citet{2020ApJ...896...63L} highlighted a decrease of the magnetic field strength in the VLBI core of \txsA{},  inferred from the core shift and variability analysis. 
They suggested that the lower value of magnetic field strength after the neutrino detection could be linked to the conversion of magnetic energy density to particle energy density. All these features observed in the parsec-scale region of \txsA{} can be framed in the scenario of this source as the emitter of \icntxs.

Moreover,  population studies carried out with VLBI data by \citet{2020ApJ...894..101P,2021ApJ...908..157P}  suggested a  correlation between bright VLBI sources and  IceCube neutrino events. In particular, they stressed the fact that the region of neutrino production lies at the base of the jet, only reachable by the high-resolution VLBI observations.
 \citet{2020ApJ...894..101P,2021ApJ...908..157P}  performed the same statistical analysis with low-resolution radio data of a sample of non-VLBI-selected sources, finding no evidence of a connection between   radio emission and neutrino events.  
 
To explore the association between blazars and neutrinos,  we  investigated 
  candidate blazar-like neutrino counterparts. In this work, we present high-resolution multi-frequency VLBI follow-ups dedicated to four high-energy neutrino events among those  occurred between 2019 and 2020. In particular, due to the considerations in favor of a connection between neutrino events and  $\gamma$-ray emission, as in the case of \txsA{}, we focused on radio counterparts of    $\gamma$-ray blazars close to the neutrino event,  most of them lying within the  90\% event error region.  Adopting this approach we attempt to determine if recurring morphological and/or evolution radio properties, as the one observed in \txsA{},  emerge in  $\gamma$-ray blazars in spatial coincident with neutrino emission.

This paper is organised as follows. We first introduce the four neutrino events and the candidate counterparts in Sect.~\ref{sample}. The VLBI observations are presented in  Sect.~\ref{observation} and results and analysis of the observations in Sect.~\ref{result}. We discuss our findings in Sect.~\ref{discussion} and conclude in Sect.~\ref{concl}. In Appendix~\ref{appendix} we briefly present results on observed sources which are less favored neutrino candidates according to our discussion. Throughout this paper we assume a $\Lambda$CDM cosmology with $H_{0} = 70$ km s$^{-1}$ Mpc$^{ -1}$, $\Omega_M = 0.3$ and  $\Omega_{\Lambda} = 0.7$.

\begin{table*}[htb]
\resizebox{18.5cm}{!} {

\centering
\begin{threeparttable}
\caption{Properties of the IceCube neutrino events analysed in this work.}
 \begin{tabular}{c c c c c c c c c}
  \hline   \hline 
IC event & Date & Alert type& RA(J2000) & DEC(J2000) &Loc. region (90\%)& Energy    & $\gamma$-ray sources &  $\gamma$-ray sources  \\
{}     &        &               &            & &         (deg$^{2}$)        & (TeV)     & (inside 90\%) &  (outside 90\%)\\
\small(1) & \small(2) &\small (3) &\small (4) &\small (5) &\small (6) & \small(7) &\small (8)&\small  (9)\\
\hline
190704A & 2019-07-04 & Bronze & $10\mathrm{h}47\mathrm{m}24.00\mathrm{s}$ & 27\deg$06' 36''$ & 20.1 & 155 & 2 & \\
200109A & 2020-01-09 & Gold   & $10\mathrm{h}57\mathrm{m}57.60\mathrm{s}$ & 11\deg$52' 12''$ & 26.6 & 375 & 2 &  1\\  
201021A & 2020-10-02 & Bronze & $17\mathrm{h}23\mathrm{m}16.80\mathrm{s}$ & 14\deg$33' 00''$ & 5.98 & 105 & 1 &  1\\
201114A & 2020-11-14 & Gold  & $07\mathrm{h}01\mathrm{m}00.00\mathrm{s}$ &  06\deg$03' 00''$ & 3.66 & 214  & 1 & \\
 \hline
 \end{tabular}
 \begin{tablenotes}[para,flushleft]
\item \textit{Notes:}
Col. 1 -  IceCube event name;
Col. 2 -  Date of the detection;
Col. 3 -  IceCube event classification;
Col. 4 - Col. 5 Neutrino best-fit position (RA, DEC);
Col. 6 -  Localization area (90\% PSF containment);
Col. 7 -  Neutrino energy;
Col. 8 -  Number of  $\gamma$-ray counterparts inside the 90\%  localization region;
Col. 9 -  Number of  $\gamma$-ray counterparts outside the 90\%  localization region.

 \end{tablenotes}
\label{tab:neutrino}
\end{threeparttable}
}
\end{table*}

\section{The IceCube neutrino events}\label{sample}

In 2016, the IceCube neutrino Observatory started a real-time alert program that releases public notifications on high-energy neutrino detections. These are collected in the GCN Circulars Archive\footnote{The acronym  GCN stands for "gamma-ray burst (GRB) Coordinates Network". Despite the  name, the archive collects alerts on all kinds of  transients (not only on GRB). The archive can be consulted at   \href{https://gcn.gsfc.nasa.gov/gcn/gcn3\_archive.html}{https://gcn.gsfc.nasa.gov/gcn/gcn3\_archive.html}}. The aim is to timely inform the astronomical community in order to obtain multi-wavelength data coincident in time and position with the neutrino arrival. Between July 2019 and  November 2020 we have performed VLBI follow-up observations of four IceCube neutrino alerts. 
 
 Considering the large uncertainties on the associations between neutrinos and astrophysical sources, and the limited sample of neutrino events collected by the  IceCube detector every year, we decided to not apply stringent criteria on which events (and relative candidate counterparts) to follow. The intent of our study is rather to present new VLBI results on blazars as possible neutrino counterparts.   
 We devoted follow-up observations to neutrino events that have arisen interest in the multi-messengers astronomy community. Most of the events have been indeed investigated with multi-wavelength follow-ups. In this context, the VLBI  results presented in this work also aim to be complementary to other-wavelength studies \citep[e.g.,][]{deMenezes:2021Fl}. 

In Table~\ref{tab:neutrino} we report basic properties of the neutrino events. According to the most recent IceCube classification, neutrino detections are divided into Gold or Bronze class. Gold alerts are announced for high-energy neutrino track events that are at least 50\% on average likely of astrophysical origin while Bronze alerts have a 30\% probability\footnote{\href{https://gcn.gsfc.nasa.gov/doc/IceCube\_High\_Energy\_Neutrino\_Track\_Alerts\_v2.pdf}{https://gcn.gsfc.nasa.gov/doc/IceCube\_High\_Energy
\_Neutrino\_Track\_Alerts\_v2.pdf}}.  We targeted two Gold and two Bronze events.  Bronze events still belong to a sample of well-reconstructed neutrino events with an average likelihood of being of cosmic origin only 20\% lower compared to the Gold sample. The two Bronze events
presented in this work have indeed drawn attention based on the possible connection with astrophysical sources, as it was the case in literature with previous detections showing similar likelihood  of astrophysical origin \citep[e.g., the IC\,41209A -- GB6\,J1040+0617 coincidence in][]{2019ApJ...880..103G}. 
 
 The energies of the neutrino events vary between  $\sim$100 TeV and  375 TeV and the arrival localisation area spans from $\sim$4 to 27 square degrees (90\% error region).
 \citet{2020ApJ...894..101P} pointed out that the 90\% error regions around the neutrinos best-fit positions published in the IceCube notifications do not take into account all the systematic errors as, for example, the ones due to the not trivial characterisation of the ice (inside which the IceCube detector operates). 
 We did not include an estimation of these systematic errors to the published error regions  (as  \citet{2020ApJ...894..101P} and others authors did).

 The  neutrino candidate counterparts  targeted by our VLBI follow-ups are the ones that have been notified in the GCN circulars and/or in the Astronomer's Telegrams\footnote{\href{https://www.astronomerstelegram.org/}{https://www.astronomerstelegram.org/}}  (ATels) dedicated to the four events. This second network collects  notices about follow-up observations (usually quickly triggered after the GCN alerts) of  transient objects.  In these follow-up campaigns the selection criteria were not uniquely established but rather defined on one-by-one case. Our final sample partly reflects this comprehensive/all-encompassing approach and is admittedly formed by an heterogeneous collection, both in neutrino events and candidate counterparts. Also, in  light of the possible  underestimation of the localisation area,   we observed  sources outside the 90\% error region when these are indicated as possible counterparts in the ATels.

 Based on the scenario of a connection between neutrinos and  $\gamma$-ray emission in blazars \citep[e.g.,][]{IceCube2018a,Kadler2016}, among the candidates reported in the ATels that we observed, we primarily analysed those sources with a gamma-ray association. For this reason, we gave priority to ATels published by 
  the \fermi{} Collaboration. The only exception was  the case of the last event, \icnD{}, for which we observed also two non-$\gamma$-ray associated sources as they are reported in the dedicated ATels.
  Each neutrino event has between   1 and 3 associated $\gamma$-ray sources, already cataloged or detected after the neutrino event with a dedicated analysis. Most of these are indeed part of the  \fermi-LAT Fourth Source Catalog  \citep[4FGL,][]{2020ApJS..247...33A} or  its  incremental  version \citep[4FGL-DR2,][]{2020ApJS..247...33A}. 
The properties of the $\gamma$-ray candidate counterparts are summarized in Table~\ref{tab:gammaray}.  The two  $\gamma$-ray sources 4FGL\,J1114.6+1225, and 4FGL\,J1728.0+1216, lie outside the 90\% error region of the \icnB{} and \icnC{} event, respectively.  The first one, 4FGL\,J1114.6+1225, was initially identified as the possible counterpart of the neutrino event by \citet{2020ATel13402....1G}.  The spatial coincidence of 4FGL\,J1728.0+1216 with \icnC{} was reported in the GCN circular 28715, sent by the IceCube Collaboration.

In Table~\ref{tab:counterpart} we report properties of the sources associated to the $\gamma$-ray ones.  In the majority of the case we used the association reported in the ATels and GCN circulars. For $\gamma$-ray sources already present in the \fermi{} catalog, we verified that the counterparts proposed in the circulars coincide with the ones in the catalog.   Only in the case   of 4FGL\,J1114.6+1225 the association is not reported in the ATEL, then we consulted the NASA/IPAC Extragalactic Database  (NED), according to which the nearest source is the infrared object \icnBone{}. We identified the  possible radio counterpart of this in the  Very Large Array (VLA) surveys (see the next Section). In these surveys, we found a detection  located at   a distance of about 2 arcmins  from the infrared position reported in the NED.

 The classification of the associated sources  listed in Table~\ref{tab:counterpart} (column 8) is retrieved from the \fermi{} catalog.  When this information is not present in the \fermi{} catalog (in the case of new, non-cataloged and/or without association  sources), it is taken from  the  NED.
In the context of blazars as neutrino-emitters, among the  $\gamma$-ray-associated sources, we analysed in detail only objects classified as BLL or FSRQ. Sources that are not confirmed blazars (although they display some blazar-like features)   are presented in  Appendix~\ref{appendix}.

 In  Table~\ref{tab:counterpart} we also present the two non-$\gamma$-ray-associated sources (\wisea{} and  \nvssC{}) which have been identified as possible neutrino counterparts in the  ATels dedicated to the \icnC{} event (see  Sect.~\ref{observation} and  Appendix~\ref{appendix}). Taking into account our focus on $\gamma$-ray-associated sources, we consider \wisea{} and  \nvssC{}  as less favored candidates because they are not associated with any $\gamma$-ray source.
However, other studies \citep[e.g.,][]{2020ApJ...894..101P,2021ApJ...908..157P} argue for a direct connection between neutrinos and VLBI cores, independently of the $\gamma$-ray emission. Therefore, we also analysed and briefly discussed these two sources in   Appendix~\ref{appendix}.   

Throughout the paper, we will refer to the radio counterpart of the sources listed in  Table~\ref{tab:counterpart} with the name reported in that table.

\noindent
\begin{table*}[htb]
\resizebox{18.5cm}{!} {
\centering
\begin{threeparttable}
\caption{Properties of the candidate $\gamma$-ray counterparts  for the neutrino events.}
 \begin{tabular}{c c c c c c c c c}
   \hline   \hline 
IC event & 4FGL/4FGL-DR2/Id. & RA(J2000) & DEC(J2000) & \small{neutrino sep.}& $F_{100\,\mathrm{ MeV}-100\, \mathrm{ GeV}}$  & $\Gamma$        & Ref. \\
        &           &           &            &  (deg)        &   \small{$\times 10^{-10}\pflux$}   &           &   &\\
\small(1) & \small(2) &\small (3) &\small (4) &\small (5) &\small (6) & \small(7) &\small (8)\\
\hline
\multirow{2}{*}[0.5em]{190704A} & J1045.3+2751 & $10\mathrm{h}$45$\mathrm{m}$22.32$\mathrm{s}$  & 27\deg$50'52.80''$ & 0.80 & 9.8$\pm$4.6  & 1.88$\pm$0.16 & (1) \\
                             {} & J1049.8+2741 & 10$\mathrm{h}$49$\mathrm{m}$50.40$\mathrm{s}$  & 27\deg$40'48.00''$ & 0.79 & 19.8$\pm$8.4 & 2.13$\pm$0.17 & (2) \\
\multirow{3}{*}[1em]{200109A}&J1103.0+1157&11$\mathrm{h}$03$\mathrm{m}$05.33$\mathrm{s}$ & 11\deg$57'55.44''$ & 1.26 & 270$\pm$19  & 2.41$\pm$ 0.03 & (2)\\
                    {}    & J1114.6+1225$^{a}$ & 11$\mathrm{h}$14$\mathrm{m}$39.36$\mathrm{s}$  & 12\deg$25'06.24''$ & 4.12 &20.2$\pm$1.1 & 2.27$\pm$0.22 &  (2)\\
                    {}    & J1055.8+1034 & 10$\mathrm{h}$55$\mathrm{m}$52.80$\mathrm{s}$  & 10\deg$34'48.0''$ & 1.38 & 16$\pm$8  & 2.06$\pm$0.18  & (3)\\
\multirow{2}{*}[0.5em]{201021A}& J1728.0+1216$^{a}$ & 17$\mathrm{h}$28$\mathrm{m}$04.85$\mathrm{s}$  & 12\deg$16'32.20''$ & 2.56 & 238$\pm$26& 2.45$\pm$0.05& (2)\\
                                & J1725.5+1312 & 17$\mathrm{h}$23$\mathrm{m}$02.40$\mathrm{s}$  & 14\deg$23'24.00''$ & 0.16 & 18$\pm$5 & 2.2$\pm$0.2 &  (4) \\
201114A &\gammasfe{}&06$\mathrm{h}$58$\mathrm{m}$33.60$\mathrm{s}$ & 06\deg$36'00.00''$ & 0.81 & 3.2$\pm$1.1 & 1.97$\pm$0.11 & (2)\\
 \hline
 \end{tabular}
 \begin{tablenotes}[para,flushleft]
\item \textit{Notes:}  
Col. 1 -  IceCube event name;
Col. 2 -  4FGL/4FGL-DR2 or identification (in case of  new sources) of candidate $\gamma$-ray counterpart;  
Col. 3 - Col. 4 - $\gamma$-ray counterpart coordinates  (RA, DEC);
Col. 5 - Angular separation between $\gamma$-ray counterpart  and the   best-fit position of the event;
Col. 6 -  Integral photon flux from 100 MeV to 100 GeV;
Col. 7 -   $\gamma$-ray  photon index.
Columns  2, 3, 4 and 7   are taken from the LAT 10-year Source Catalog 4FGL/4FGL-DR2. 
Col. 8. \fermi{}  catalog  reference: (1) = 4FGL-DR2 and (2) = 4FGL/4FGL-DR2; ATels reference: (3) = \citet{2020ATel13402....1G},  (4) = \citet{2020ATel14111....1B}.
The sources marked with $^{a}$ lie   outside the 90\% neutrino localization region.
 \end{tablenotes}
\label{tab:gammaray}
\end{threeparttable}
}
\end{table*}

\begin{table*}[htb]
\resizebox{18.5cm}{!} {
\centering
\begin{threeparttable}
\caption{Information on other-wavelength   associations with the candidate   counterparts  of neutrino events. }
 \begin{tabular}{c c c c c c c c c}
  \hline   \hline 
    \noalign{\smallskip}
IC event &  4FGL/4FGL-DR2/Id. & \multicolumn{6}{c}{Counterpart}     \\
       &           &  Name     & RA(J2000)         &     DEC(J2000)       &     {($'$) from $\nu$} &  {($'$) from $\gamma$} &  {Class}    & {$z$} \\
\small(1) & \small(2) &\small (3) &\small (4) &\small (5) &\small (6) & \small(7) & \small(8) & \small(9) \\
\hline
\multirow{3}{*}[1em]{190704A} & \gammajtff{}   &  {\whsp{}} & 10$\mathrm{h}$45$\mathrm{m}$16.30$\mathrm{s}$ & 27\deg$51'33.46''$ & 53 & 1.5 &  {BLL} &  {1.914}  \\
           {}                 & J1049.8+2741 &  {\nvssA{}}$^{b}$ & 10$\mathrm{h}$49$\mathrm{m}$38.80$\mathrm{s}$ & 27\deg$42'13.00''$ & 47 &  2.9 &  {G}   &  {0.144} \\
\multirow{3}{*}[1em]{200109A} & J1103.0+1157 &  {\txsB{}}          & 11$\mathrm{h}$03$\mathrm{m}$03.53$\mathrm{s}$ & 11\deg$58'16.62''$ &  75 & 0.6 &  {FSRQ} &  {0.91} \\
            {}     & J1114.6+1225 &  {\icnBone{}$^{b}$-A}  & 11$\mathrm{h}$14$\mathrm{m}$37.02$\mathrm{s}$ & 12\deg$27'13.12''$ & 247 & 2.2 &  {IrS} & - \\
                &  &    {\icnBone{}$^{b}$-B} & 11$\mathrm{h}$14$\mathrm{m}$29.76$\mathrm{s}$ & 12\deg$28'03.40''$ & 245 &  3.8 &   & -   \\
            {}     & J1055.8+1034 &  {\icnBtwo{}}$^{b}$  & 10$\mathrm{h}$56$\mathrm{m}$47.79$\mathrm{s}$ & 10\deg$30'28.10''$ & 84 & 14.2 & {IrS}   & - \\
\multirow{2}{*}[0.5em]{201021A}& J1728.0+1216      &  {\pks{}} & 17$\mathrm{h}$28$\mathrm{m}$07.05$\mathrm{s}$ & 12\deg$15'39.49''$ & 154 & 1.0 & {FSRQ} &  {0.568} \\
                               & J1725.5+1312  &  {\rxs{}}$^{b}$ & 17$\mathrm{h}$23$\mathrm{m}$14.12$\mathrm{s}$ & 14\deg$21'00.62''$ & 12 & 3.7 &  {XrayS} & - \\
\multirow{3}{*}[1em]{201114A} & J0658.6+0636 &  {\nvssB{}}    & 06$\mathrm{h}$58$\mathrm{m}$45.02$\mathrm{s}$ & 06\deg$37'11.49''$& 48 & 3.0   &  {BCU} & - \\
 {}                 & -  & {\wisea{}}                    & 06$\mathrm{h}$56$\mathrm{m}$33.43$\mathrm{s}$ & 05\deg$39'22.87''$ & 70 & -&  {BC}  & - \\
 {}                 & -  &  {\nvssC{}}                       & 06$\mathrm{h}$59$\mathrm{m}$18.00$\mathrm{s}$ & 08\deg$13'30.95''$ & 132 &- & {BC}  & - \\
 \noalign{\smallskip}
 \hline
 \end{tabular}
 \begin{tablenotes}[para,flushleft]
\item \textit{Notes:}  
Col. 1 -  IceCube event name;
Col. 2 -  4FGL/4FGL-DR2 or identification (in case of  new sources) of candidate $\gamma$-ray counterpart;  
Col. 3 -  Name of the source associated with the $\gamma$-ray candidate neutrinos counterpart  and of the candidate neutrinos counterparts without a $\gamma$-ray association;
Col. 4 - Col. 5 - Radio coordinates (RA, DEC, inferred from the   VLBI  observations analysed in this work) of the radio sources associated with the objects indicated in column 3.   There are two possible radio counterparts associated with  \icnBone{}, indicated with A and B.  
Col. 6 -  Angular separation of the associated radio source from   the neutrino  ($\nu$) best-fit position and  Col. 7 -  from  the  $\gamma$-ray source best-fit position;
Col. 8 -  Classification of the associated source  from the \fermi{} catalog or from the  NED. The latter catalog is consulted when the $\gamma$-ray source are non-associated in the \fermi{} catalog. These sources are marked with $^{b}$.
Col. 9 - Spectroscopic redshift of the associated source. WISEAJ065633.43+053922.7 and NVSS J065916+055252  have no $\gamma$-ray  counterpart reported. 
(BLL = BL Lac; G = galaxy object; FSRQ = Flat Spectrum Radio Quasar; IrS = Infrared Source; XrayS = X-ray Source; BCU = Blazar Candidate or Unknown in the 4FGL classification; BC = Blazar Candidate according to the VOU\_Blazar tool \citep[the tool is described in][]{2020A&C....3000350C}, the BC classification is  reported in \citet{2020ATel14225....1G}).
 \end{tablenotes}
\label{tab:counterpart}
\end{threeparttable}
}
\end{table*}


\section{Observations and data reduction}\label{observation}

In order to investigate the radio structures and variability of the candidates, we analysed both VLBI datasets taken after the neutrino detection and archival data. Carried out on longer integration time and at different observing frequencies,  our new multi-frequency VLBI observations allow us to produce  higher quality and higher fidelity images with respect to the available archival data of the targets. Moreover, some of   them have never been observed  at VLBI resolution before.

\subsection{New VLBI data}

The  VLBI follow-up observations of the four events have been performed  with the Very Long Baseline Array (VLBA), the European VLBI Network (EVN), and the e-MERLIN array.   Details about the observations are listed in Table~\ref{tab:obs}. Blazar sources are expected to experience variability in the radio emission at the GHz regime   on timescales of several weeks \citep[e.g.,][]{2013MNRAS.428.2418O}. Therefore our VLBI observations were carried out with time gaps from the neutrino detection in the range between few days and one month. 
The observation frequencies, from 1.5  GHz to 23.5 GHz,  are selected to optimise the balance between sensitivity and resolution of the  jet structure on different  mas-scales.  The total bandwidth ranges between 32 MHz, 64 MHz and 128 MHz (until 512 MHz for the e-MERLIN array, see below).
All the sources have been observed at  least at two observing frequencies. 
Only in the case of \pks{} and \rxs{} we obtained a single-frequency observation. The  sensitivity levels of the images  are  between 20  {$\mathrm{\mu Jy\,beam^{-1}}$} and 100  {$\mathrm{\mu Jy\,beam^{-1}}$}. The  restoring beam sizes go from 0.8 mas to 12 mas in VLBA and EVN images and from  35 mas to 84 mas in e-MERLIN images.  Antennas participating in the observations are reported in Table~\ref{tab:obs}. 

In processing all the raw VLBI data, we applied the standard approach of VLBI data reduction  described in the NRAO Astronomical Image Processing System \citep[AIPS,][]{2003ASSL..285..109G} cookbook\footnote{\href{http://www.aips.nrao.edu/cook.html}{http://www.aips.nrao.edu/cook.html}}, which includes  visibility amplitude calibration, bandpass calibration and phase calibration. This calibration procedure was  carried out either with the  AIPS software package or with the Common Astronomy Software Applications 
 \citep[CASA,][]{2007ASPC..376..127M}. 
After these steps, we exported the single-source visibilities   from AIPS or CASA and imported them into the DIFMAP software  \citep{1994BAAS...26..987S} for    self-calibration  and imaging. For the imaging, the DIFMAP software follows  the  CLEAN approach which includes fitting Delta functions to the visibility data \citep{1974A&AS...15..417H}.

\begin{itemize}
    \item \textbf{\icnA{}} The follow-up observations of the event were focused on the two  candidates listed in Table~\ref{tab:counterpart}, \nvssA{}  and \whsp{}.  The candidates  were observed less than one month after the neutrino detection, on 2019 July 18 and 22, for a total of 3-hr of observation at 1.5 GHz, 4.4 GHz, 7.6 GHz and 8.4 GHz.  We separately  calibrated the datasets of the two days  and then concatenated them in one.  After a first inspection of the observation, we decided to dedicate a second deeper (4-hr) observation   only to the blazar-like source,  \whsp{} (see Sect. \ref{result}). This second observation  was performed  on 2020 January 17, using the wide C-band with two intermediate frequency bands (IFs, i.e. sub-bands) centered at 4.7 GHz and two IFs centered at 7.7 GHz. The calibrated  data were  split  into two halves  with one half containing the first two IFs  and the other half containing the last two IFs.  
     Due to the  faintness of the targets  and the uncertainties on their  coordinates, the  observations   were performed adopting the phase-referencing mode.   The phase-calibrator is a nearby bright FSRQ, J1037+2834 (B1034--2551, taken from the  VLBA Calibrator Survey, VCS\footnote{\href{http://astrogeo.org/vcs/}{http://astrogeo.org/vcs/}}). 
   It  is outside the 90\% neutrino error region and it is not associated with  $\gamma$-ray sources.

    \item \textbf{\icnB{}} The VLBA and EVN follow-up observations of the event included the  four targets reported in Table~\ref{tab:counterpart}. In particular, in spatial coincidence with 4FGL\,J1114.6+1225 (associated with \icnBone{}), there are two possible radio counterparts. In the course of the manuscript we will refer  to them using the name \icnBone{} followed by  the suffixes A and B.  The observation  was carried out at 8.4 GHz and 23.5 GHz with the VLBA and at 4.9 GHz with the EVN. The position of the candidate \txsB{} is constrained with a precision of the order of 0.1 mas \citep[coordinates from VCS]{2020ATel13397....1K}. Also, it is bright enough to allow for fringe-fitting in the calibration procedures. Both EVN and VLBA data of the other two candidates, \icnBone{} and \icnBtwo{}, have been calibrated using \txsB{}. 
    
    \item \textbf{\icnC{}} About 14-hr of e-MERLIN observing time were spent on the two possible \icnC{} counterparts: \rxs{} and \pks{}. The  e-MERLIN interferometer provides different  total bandwidth and angular resolution than VLBA and EVN. The  e-MERLIN observations were carried out  at 5.1 GHz with a bandwidth of 512 MHz. The  angular resolution of e-MERLIN, of the order of 30--80 mas, corresponds to larger linear scales with respect to VLBA and EVN ones. 
    
    Being a bright source, \pks{}  was used as phase-reference calibrator.  
    
    The e-MERLIN data reduction was performed using the e-MERLIN CASA Pipeline v1.1.19 \citep{2021ascl.soft09006M}.

     \item \textbf{\icnD{}} The VLBA and EVN observations of \icnD{} candidate counterparts are described in  Table~\ref{tab:counterpart}.  Since the most favored candidate (see Sect.~\ref{result}), \nvssB{}, was known to be a faint radio source from Radio Fundamental Catalogue (RFC, described below)  data, both VLBA and EVN  observations were carried out in phase-referencing mode. \nvssC{}, also identified as possible radio counterpart of the neutrino event  \citep{2020ATel14225....1G},  is the phase calibrator for \nvssB{}, with an offset of 1.6 degrees.  The other candidate reported in \citet{2020ATel14225....1G}, \wisea{}, was included in the observation schedule as check-source of the dataset. 
     
\end{itemize}

\begin{table*}[htb]
\centering
\small
\begin{threeparttable}
\caption{Summary of VLBI observations.}
 \begin{tabular}{c c c c c c c c}
 \hline \hline
IC event & Target & Date      & Code & Array & $\nu$            & $t_{\mathrm{obs}}$       &  Antennas \\ 
{}     &           &      &   &    & $\mathrm{(GHz)}$ & $\mathrm{(min)}$& \\
\small(1) & \small(2) &\small (3) &\small (4) &\small (5) &\small (6) & \small(7) &\small (8)\\
\hline 
 190704A & \whsp{}& 2019-07-18/22 & BG261 & VLBA & 1.5 & 14+13 & 9 ($-$Sc)\\

        &     &          &       &      & 4.4 & 9+8   &  \\
        &     &          &       &      & 7.6 & 9+8   &  \\
        &     &          &       &      & 8.4 & 17+18 &  \\
       & & 2020-01-17    & BA133 & VLBA & 4.7 & 175   & 9 ($-$Kp) \\
{}      &      &         &       &      & 7.6 & 175   &   \\
         &  \nvssA{}&  2019-07-18/22 & BG261 & VLBA & 1.5 & 13+14 &  \\
        &      &         &       &      & 4.4 & 8+9   & \\
        &      &         &       &      & 7.6 & 9+9   &\\
\hline
 200109A & \txsB{} &  2020-02-29   & RG011  & EVN & 4.9 &  525   & 13\\
         & &  2020- 02-04  & BG263  & VLBA & 8.4 &  60  & 10 \\
        &  &               &        &     & 23.5 &  270  & \\
 &\icnBone{}-A &  2020-02-29 & RG011  & EVN & 4.9 &  34  &   \\
 &\icnBone{}-B &          &       &      & 4.9 &  34  & \\
 &\icnBone{}  &   2020- 02-04  & BG263  & VLBA & 8.4  &  14  &  \\
            &     &     &        &     &    23.5 &  36  &  \\
&\icnBtwo{} &  2020-02-29 & RG011  & EVN & 4.9 &  33   &  \\
             &  &  2020- 02-04  & BG263  & VLBA & 8.4&  14   &  \\
            &     &     &        &     &   23.5  &   36  &  \\
\hline
201021A & \rxs{}& 2020-11-05  & DD10006&e-MERLIN& 5.1 & 486 & 6 \\
 & \pks{}&             &        &        &     & 155 &      \\
\hline
 201114A & \nvssB{}& 2020-12-01/02& EG108 & EVN & 4.9& 328  & 16 \\
     &   & 2020-12-06   & BG264A & VLBA & 8.4 & 72 & 9 ($-$Hn) \\
      &  &              &        &      & 23.5& 198 &  \\
 &\wisea{}& 2020-12-01/02 &  EG108 & EVN  & 4.9   &  20   &   \\
        & & 2020-12-06   & BG264A & VLBA & 8.4 &  18   &  \\
        &     &         &        &      & 23.5&  46   &  \\
&\nvssC{}& 2020-12-01/02 &  EG108 & EVN &  4.9    & 252    &    \\
        & &  2020-12-06   & BG264A & VLBA  & 8.4 &  43  &  \\
        &     &         &        &      & 23.5&  139  &  \\
        \hline
 \end{tabular}
\begin{tablenotes}[para,flushleft]
\item \textit{Notes:} 
Col. 1 -  IceCube event name;
 Col. 2 -  Candidate neutrino counterpart; 
Col. 3 -  Date of observation;
Col. 4 -  Project code;
Col. 5 -  Instrument;
Col. 6 -  Observation frequency in GHz; 
Col. 7 -  On-source time in minute;
Col. 8 -  Number of antennas used in the observations. We report in brackets which antenna was not operating during VLBA experiments. VLBA telescopes are: Saint Croix (Sc), Kitt Peak (Kp),  Hancock (Hn),  Mauna Kea, Brewster, Owens Valley,  Pie Town,  Los Alamos, Fort Davis, North Liberty. The EVN telescopes partecipating in the observations are:  Jodrell Bank (Jb); Onsala (O8); Tianma  (T6); Nanshan (Ur); Torun (Tr); Yebes (Ys); Svetloe (Sv); Zelenchukskaya (Zc); Badary (Bd); Irbene (Ir);  Westerbork (Wb); Effelsberg  (Ef);  Medicina (Mc); Noto (Nt);  Hartebeesthoek (Hn).  In particular, the RG011 project is performed with Jb, O8, T6, Ur, Tr, Ys, Sv, Zc, Bd, Ir, Wb, Ef, Hh; the EG108 project is performed with Jb, Wb, Ef, Mc, Nt, O8, T6, Tr, Ys, Hn, Sv, Zc, Bd, Ir, and the e-MERLIN stations Cambridge (Cm); Darnhall (Da);  Defford (De);  Knockin (Kn);  Pickmere (Pi).
\end{tablenotes}
	\label{tab:obs}
\end{threeparttable}
\end{table*}

\subsection{Archival data}

 In addition to the proprietary data, we  analysed archival VLBI data and publicly available survey larger-scales data  in order to have a more complete picture of the characteristics of the sources. Most of the archival VLBI data are retrieved from  the RFC, which contains raw and calibrated data and images of thousands of sources. The RFC collects datasets   of  observations   devoted to calibrators monitoring or to  astrometry experiments.  Archival  data of \pks{} used in this work are taken from   the  Monitoring Of Jets in Active galactic nuclei with VLBA Experiments \citep[MOJAVE,][]{2018ApJS..234...12L}\footnote{\href{http://www.physics.purdue.edu/astro/MOJAVE/index.html}{http://www.physics.purdue.edu/astro/MOJAVE/index.html}}. Therefore we will explicitly refer to MOJAVE data in the  case of \pks{}. 
 We re-imaged the available  calibrated data from the  RFC and MOJAVE with the DIFMAP software. The properties of the RFC and MOJAVE images   are summarized in Table~\ref{tab:rfc}. The short observing duration of archival observations  (of the order of minutes)  results in a poorly sampled $uv$-plane.

The arcsecond-scale extended emission of the sources was studied using  VLA survey observations. A comparison between  our VLBI data and information taken from VLA surveys   helps to determine the nature of the targets, from  mas- to arcsecond-scales. In particular, we retrieved the  images  of the targets from the NRAO VLA Sky Survey \citep[NVSS;][]{1998AJ....115.1693C}, the  Faint Images of the Radio Sky at Twenty-cm \citep[FIRST;][]{1994ASPC...61..165B} survey and the VLA Sky Survey \citep[VLASS;][]{2020PASP..132c5001L}\footnote{The VLASS images are taken from the Canadian Initiative for Radio Astronomy Data Analysis (CIRADA) catalog \citep[www.cirada.ca/catalogues;][]{2020RNAAS...4..175G}.}. The VLASS is an ongoing project and the final catalog is not released yet. The VLASS images  taken from the CIRADA catalog  are produced using a simple imaging algorithm and no self-calibration is applied. This   limits the accuracy of the results that we inferred from these images. The surveys are carried out at 1.4 GHz (NVSS and FIRST) and 3 GHz (VLASS).  NVSS and FIRST images   are characterised by a beam of   45$''$ $\times$ 45$''$ and 5.4$''$ $\times$ 5.4$''$, respectively, while the VLASS has a  resolution of about 3$''$ $\times$ 2$''$. For some sources, the VLASS  has two runs, denoted as 1.1 and 1.2 in Table~\ref{tab:nvss-first}. This Table reports the  properties of all the archival images.

\begin{table*}[htb]
\centering
\begin{threeparttable}
\caption{Properties of RFC and MOJAVE observations.}
 \begin{tabular}{l c c c c c c c c}
  \hline   \hline 
 Source  &  Date & $\nu$            &$S_{\mathrm{peak}}$                 &{$S_{\mathrm{int}}$}       &rms                           &Beam \\
         &      &  \small{{$\mathrm{(GHz)}$}}    &\small{{$\mathrm{(mJy\,beam^{-1})}$}} & \small{{$\mathrm{(mJy)}$}} & \small{{$\mathrm{(mJy\,beam^{-1})}$}} & \small ({mas$\times$mas,\deg}) \\
\small(1) & \small(2) &\small (3) &\small (4) &\small (5) &\small (6) & \small(7) \\
\hline
\multicolumn{5}{l}{\textbf{200109A}} \\
\multirow{3}{*}[1em]{\txsB{}}&   2004-04-30 & 2.3 & 268$\pm$27 & 310$\pm$31  &  0.6 & 7.4$\times$3.2, $-$0.1\\
                              &    2004-04-30 & 8.6 & 279$\pm$28 & 311$\pm$31  & 0.7 & 2.0$\times$0.9, 1.1\\
                            &   2007-08-01 & 8.4 &   353$\pm$35 & 403$\pm$41  & 0.4 & 2.2$\times$1.2, 35.1\\
                           &     2012-02-20 & 8.4 & 105$\pm$11 & 150$\pm$15  &  0.3  & 2.0$\times$0.9, 8.0\\ 
\hline
\multicolumn{5}{l}{\textbf{201021A}} \\
\multirow{3}{*}[1em]{\pks{}}&  2018-10-06  & 15.3 & 579$\pm$58 &   622$\pm$62  & 0.1   & 1.2$\times$0.6, 6.7 \\
                              &    2019-07-19 & 15.3 & 495$\pm$50 & 509$\pm$51  & 0.1  & 1.8$\times$0.6, $-$18.1\\
                           & 2020-05-25 & 15.3 &    460$\pm$46 &   470$\pm$47 & 0.1 &  1.8$\times$0.7, $-$21.6\\
                           &     2020-10-21 & 15.3 &   530$\pm$53 &   555$\pm$56 &   0.09 & 1.1$\times$0.5, $-$4.6\\ 
                        &     2020-12-01 & 15.3 &  638$\pm$64  &  657$\pm$66   &   0.08  & 1.2$\times$0.6, $-$4.8\\ 

\hline
\multicolumn{5}{l}{\textbf{201114A}}\\
\multirow{5}{*}[2.3em]{\nvssB{}}  & 2013-04-08/09$^{*}$ & 4.3 & 18.8$\pm$1.8  & 22.5$\pm$2.3& 0.2  & 4.6$\times$1.9, $-$7.3 \\
                &  2013-04-08/09$^{*}$ & 7.6 & 17.9$\pm$1.8 & 22.7$\pm$2.3 & 0.2 & 2.8$\times$1.1, $-$12.2\\
                &  2013-10-19 &  7.6 & 10.7$\pm$1.1  & 15.2$\pm$1.5 & 0.09 & 2.2$\times$1.3, $-$3.3 \\
                \hline

 \end{tabular}
 \begin{tablenotes}[para,flushleft]
\item \textit{Notes:} 
Col. 1 -  Candidate neutrino counterpart;
Col. 2 -  Date of the observation;
Col. 3 -  Observation Frequency in GHz;
Col. 4 -  Peak brightness  in mJy~beam$^{-1}$;
Col. 5 -  Integrated  flux density in   mJy; 
Col. 6 -  1-$\sigma$ noise level of the image in $\mathrm{mJy\,beam^{-1}}$; 
Col. 7 -  Major axis (in mas), minor axis (in mas), and position angle (in degrees, measured from North to East)  of the restoring beam.
The parameters are referred to  natural weighting images. Observations marked with $^{*}$ were originally separated datasets that we concatenated  in one dataset, being these observations made in close days.
\end{tablenotes}
\label{tab:rfc}
\end{threeparttable}
\end{table*}

\begin{table*}[htb]
\centering
\begin{threeparttable}
\caption{Properties of NVSS, FIRST and VLASS observations.}
 \begin{tabular}{l c  c c c c }
  \hline   \hline 
   \noalign{\smallskip}
   
 Source &   Survey & $\nu$ & Date    &  $S_{\mathrm{peak}}$ & $S_{\mathrm{int}}$ \\
        &      &  (GHz)  &        & {$\mathrm{(mJy\,beam^{-1})}$} &  {$\mathrm{(mJy)}$}  \\ 
\small(1) & \small(2) &\small (3) &\small (4) &\small (5) &\small (6)  \\
\hline
\textbf{190704A} &    & & \\
 \whsp{} & NVSS & 1.4 & 1994-01-11   & 2.7$\pm$0.3  & 2.7$\pm$0.4 \\
       & FIRST &  1.4 & 1995-11-04 &  3.5$\pm$0.4  &  3.5$\pm$0.4  \\
        & VLASS 1.2 & 3 & 2019-06-08 & 2.5$\pm$0.4  & 2.5$\pm$0.4  \\
\hline
 \nvssA{} & NVSS & 1.4 & 1994-01-11   &  18.0$\pm$2.0 &  20.0$\pm$2.2 \\
        & FIRST & 1.4 & 1995-11-04 &   9.0$\pm$1.0 &  15.6$\pm$1.6\\
        & VLASS 1.2 & 3 & 2019-06-08 & 7.1$\pm$1.1  & 11.6$\pm$1.7  \\
         \hline
\textbf{200109A}  &          &     & \\
 \txsB{} & NVSS & 1.4 & 1995-02-27  & 251$\pm$25  & 264$\pm$27 \\
               & FIRST & 1.4 & 2000-01-15 &   274$\pm$27  & 300$\pm$30 \\
               & VLASS 1.1 & 3 & 2017-11-22 & 282$\pm$28 & 308$\pm$31  \\
             & VLASS 1.2 & 3 & 2020-07-21 & 314$\pm$31 &  344$\pm$35 \\
             \hline
 \icnBone{}-A& NVSS & 1.4 & 1995-02-27  &  2.5$\pm$0.2   &  3.1$\pm$0.3  \\
             & FIRST & 1.4 & 1995-12-15 &   3.5$\pm$0.2 &  2.9$\pm$0.2 \\
             & VLASS 1.1 & 3 &2017-12-28 &  3.9$\pm$0.6  &  4.1$\pm$0.5   \\
             & VLASS 1.2 & 3 & 2020-08-18 &  4.0$\pm$0.4  &   3.8$\pm$0.4  \\
 \icnBone{}-B & NVSS & 1.4 & 1995-02-27  &  2.7$\pm$0.2  &  3.6$\pm$0.3 \\
             & FIRST & 1.4 & 1995-12-15 &    2.7$\pm$0.2 &   3.1$\pm$0.3\\
             & VLASS 1.1 & 3 & 2017-12-28 &   2.5$\pm$0.3   &  2.8$\pm$0.3    \\
             & VLASS 1.2 & 3 & 2020-08-18 &  2.6$\pm$0.3   &   2.9$\pm$0.4  \\
    \hline
 \icnBtwo{}& NVSS & 1.4 & 1995-02-27  &  343$\pm$17 &  356$\pm$18   \\
             & FIRST & 1.4 & 2000-01-15 &   333$\pm$17 &   347$\pm$18\\
             & VLASS 1.1 & 3 & 2017-11-21 &    208$\pm$21    &   204$\pm$20    \\
             & VLASS 1.2 & 3 & 2020-07-21 &   193$\pm$19    &    192$\pm$19  \\
    \hline
\textbf{201021A} &          &     &  \\
 \rxs{} & NVSS & 1.4 & 1995-02-27  &   \multicolumn{2}{c}{<0.6$^{*}$} \\
                 & VLASS 1.2 & 3 & 2019-03-30 & 1.2$\pm$0.3  & 0.9$\pm$0.1 \\
                 \hline
 \pks{} & NVSS & 1.4 & 1995-02-27 &  335$\pm$34   &   348$\pm$35 \\
                & VLASS 1.2 & 3 & 2019-03-30 & 353$\pm$35  & 360$\pm$36    \\
\hline
\textbf{201114A} &          &     &  \\ 
 \nvssB{}  & NVSS & 1.4 & 1993-11-15   & 24.3$\pm$2.5    & 23.8$\pm$2.5 \\ 
                &VLASS 1.1  & 3 & 2017-09-15   & 19.1$\pm$1.9 & 19.7$\pm$2.1  \\
           &VLASS 1.2 & 3 & 2020-08-09     & 14.6$\pm$1.5 & 17.6$\pm$1.8 \\
           \hline
 \wisea{}-A & NVSS & 1.4 & 1993-11-15 &   53.7$\pm$2.7  & 61.5$\pm$3.2    \\
             & VLASS 1.1 & 3 & 2017-09-15   & 29.2$\pm$3.0 & 36.0$\pm$3.6  \\
             & VLASS 1.2 & 3 & 2020-08-09     & 31.2$\pm$3.2 & 31.0$\pm$3.1\\
 \wisea{}-B & NVSS & 1.4 & 1993-11-15 & 140.4$\pm$7.0 & 160.3$\pm$8.1   \\
             & VLASS 1.1 & 3 & 2017-09-15   & 47.5$\pm$4.8 & 78.0$\pm$7.8  \\
             & VLASS 1.2& 3 & 2020-08-09     & 46.2$\pm$4.6 & 61.8$\pm$6.2\\
           \hline
 \nvssC{} & NVSS & 1.4 & 1993-11-15  &  896$\pm$90  & 935$\pm$94     \\
                  &VLASS 1.1& 3 &  2017-09-15  & 723$\pm$73  & 823$\pm$83 \\
                  & VLASS 1.2 & 3 & 2020-09-21  & 829$\pm$83  & 910$\pm$91     \\
                  \hline
 \noalign{\smallskip}
 \end{tabular}
 \begin{tablenotes}[para,flushleft]
\item \textit{Notes:} 
Col. 1 -  Candidate neutrino counterpart; Col. 2 -  Survey;
Col. 3 -  Observation Frequency in GHz;
 Col. 4 -  Date of observation; 
 Col. 5 -  Peak brightness in $\mathrm{mJy\,beam^{-1}}$ and Col. 6 -   Integrated   flux density in mJy.  
$^{*}$ rms=0.2 {$\mathrm{mJy\,beam^{-1}}$}.
\end{tablenotes}
\label{tab:nvss-first}
\end{threeparttable}
\end{table*}


\section{Analysis and results}\label{result}

In Table~\ref{tab:obs_res} we list image parameters of the new VLBI data,  using  natural weighting.   We measured the peak brightness, $S_\mathrm{peak}$, and the integrated flux density,  $S_\mathrm{int}$,   of each target at each frequency. The latter is extracted from a polygonal area on the targets images using the VIEWER tool of the CASA software. The area of extraction is the one  above 3 times the root-mean-square (rms) contour levels of the images. The rms is  measured within an off-source region on the image plane. The uncertainties on $S_{\nu}$  are given by:

\begin{equation} 
\label{eq:errorfluxradio}
\sigma_{S_{\nu}} = \sqrt{(\mathrm{rms}\times \sqrt{N_{\mathrm{beam}}})^2 + \sigma_{\mathrm{cal}}^2}
\end{equation}

where $N_{\mathrm{beam}}$ is the number of beams of the area in which $S_\nu$ is extracted. 
The error in the calibration procedure, $\sigma_{\mathrm{cal}}$, is defined as:   $\sigma_{\mathrm{cal}}=\xi \times S_{\nu}$, in which we assume $\xi$   of the order of  10\% for VLBA, EVN and VLASS data and of the order  of  5\% for e-MERLIN, NVSS and FIRST data. 
When the source is unresolved, we fitted the emission with a 2-dimensional  Gaussian function with the \texttt{imfit} task in CASA.  In this case, the uncertainties on the flux densities are estimated with the sum of squares of the  fit error plus the  calibration error.  At the time of writing, we are aware of an issue with VLBA data taken starting from the first half of the 2019\footnote{\href{https://science.nrao.edu/enews/14.4/index.shtml\#vlba\_flux}{https://science.nrao.edu/enews/14.4/index.shtml\#vlba\_flux}}. 
The    effects on the VLBA flux density scale of those data are currently under investigation. Our VLBA data at 23.5 GHz could be affected by this issue. While no standard procedure has been so far indicated by the National Radio Astronomy Observatory (NRAO) team, the helpdesk has suggested a test the results of which reassure us about applied analysis procedure.  However, in this Section  and in Sect.~\ref{discussion} we briefly discuss the presence and the implications of additional errors in the flux density scales used.

 The  compactness of the sources  can result in synchrotron self-absorption  affecting the lower frequencies. Assuming $S\propto \nu^\alpha$, with  $\alpha$ being the spectral index, we fit two power-laws  for couples of adjacent frequencies to calculate the spectral index pattern over the sampled frequencies. 
 The uncertainty of $\alpha$       is calculated based on  the error propagation equation, as:

\begin{equation}\notag
\alpha \pm \Delta \alpha= \frac{ \ln \left( \frac{S_{1}}{S_{2}}\right)}{\ln \left(\frac{\nu_1}{\nu_2}\right)} 
\pm \left| \frac{1}{\ln\left(\frac{\nu_2}{\nu_1}\right)}\right|
\sqrt{
\left( \frac{\sigma_{S_1}}{S_1} \right)^2 +
\left( \frac{ \sigma_{S_2}}{S_2} \right)^2 
.}
\end{equation}

Before calculating the spectral index, both images at each frequency couple had been restored with the same $uv$-range,   pixel size and restoring beam size and shape. We  used the portion of $uv$-range   covered by the   observations at the two frequencies. 
Measurements of the spectral index are reported in Table~\ref{tab:ind-spec}.

Following the case of \txsA{} which significantly increased its radio emission during the neutrino event \citep{2019MNRAS.483L..42K,2020A&A...633L...1R,2020ApJ...896...63L}, we searched for a similar behaviour in our sources.    
To  quantitatively estimate the flux density variability for sources  for which archival VLBI observations   are available, we adopted the method used by \citet{1992ApJ...399...16A}. This consists in calculating the variability index, $V$,  with: 

\begin{equation}\label{eq:var}
V = \frac{(S_{\mathrm{max}} - \sigma_{S_{\mathrm{max}}})-(S_{\mathrm{min}} + \sigma_{S_{\mathrm{min}}})}{(S_{\mathrm{max}} - \sigma_{S_{\mathrm{max}}})+(S_{\mathrm{min}} + \sigma_{S_{\mathrm{min}}}),}
\end{equation}

where $S_{\mathrm{max}}$ and $S_{\mathrm{min}}$ are the integrated flux densities  of the higher and  lower state and  $\sigma_{S_{\mathrm{max}}}$ and  $\sigma_{S_{\mathrm{min}}}$ are the associated uncertainties. According to  \citet{1992ApJ...399...16A}, a variability  of the order of   10\%, which corresponds to $V \geq 0.1$, indicates a significant change    in the flux density of the source.  Indeed, from  the flux density data reported in \citet{2020A&A...633L...1R},   we obtain $V\sim 0.1$ in \txsA{}  over six months.   

For those sources with a known redshift, $z$, we computed a radio luminosity based on the formula:

\begin{equation}\notag
L_{\nu} = 4\pi S_{\nu} d_{L}^2 (1+z)^{\alpha-1}
\end{equation}

where $d_{L}$ is the  luminosity distance of the source. 
The luminosity reported in Table~\ref{tab:lum} are measured at around  1.4 GHz and at about 5 GHz. The spectral index  $\alpha$, adopted for the luminosity computation, is estimated between   1.4 GHz and 3 GHz data from the not simultaneous (separated by $\sim$20-25 years)  NVSS and VLASS.  

We characterised the jet emission of extended sources by fitting a two-dimensional Gaussian function with the modelfit routine in the DIFMAP software.  In this procedure, we fit the visibilities data with Gaussian model components.  A good fit is achieved when $\sigma_{\text{res}} = \sigma_{\text{cln}} \pm 10 \%$, where $\sigma_{\text{res}}$ and $\sigma_{\text{cln}}$ are respectively the rms noise level of the residual and the cleaned image. 
This ensures that the model components well describe the jet structure.   We set the uncertainty associated with each component flux density as  10\% of the flux density of the component itself. The precision associated with polar coordinates of the components, that is the radius and the position angle (P.A.),   depends on the dimensions and the orientation of the image restoring beam. We assume that the component center lies within an ellipse oriented as the beam, and with the major and minor axis equal to the 10\% of the beam ones. In some cases, it was necessary to fix the axis ratio and/or the position of the components to obtain reliable fit (see, e.g., Table~\ref{tab:mf_txs} in Appendix~\ref{modelfit}).

\begin{table*}[htb]
\centering
\small
\begin{threeparttable}
\caption{Imaging parameters of VLBI observations.}
 \begin{tabular}{l c c c c c c c }
 \hline \hline
    \noalign{\smallskip}
Source & Date       & Code & $\nu$            &$S_{\mathrm{peak}}$                 &{$S_{\mathrm{int}}$}       &rms                           &Beam\\ 
{}    &            &      & $\mathrm{(GHz)}$ &  $\mathrm{(mJy\,beam^{-1})}$&{$\mathrm{(mJy)}$}&{$\mathrm{(\mu Jy\,beam^{-1})}$} &({mas$\times$mas,\deg})\\
\small(1) & \small(2) &\small (3) &\small (4) &\small (5) &\small (6) & \small(7) &\small (8)\\
\hline 
  \textbf{190704A}\\
\whsp{} & 2019-07-18/22 & BG261 & 1.5 & 3.5$\pm$0.3 & 3.8$\pm$0.4 & 95 & 11.6$\times$6.1, $-$18.4\\
{}          &            &        &  4.4 &3.3$\pm$0.4  & 3.7$\pm$0.5 & 90 & 4.3$\times$2.2, $-$24.5\\
{}          &            &        &  7.6 & 3.8$\pm$0.4 & 3.7$\pm$0.5 & 138 & 2.4$\times$1.3, $-$20.7\\
{}          &            &        &  8.4 & 3.4$\pm$0.3 & 3.5$\pm$0.4 & 47 & 2.1$\times$1.1, $-$22.8\\
{}          & 2020-01-17 & BA133  &  4.7 & 2.5$\pm$0.2 & 2.7$\pm$0.3 & 27 & 3.5$\times$1.4, $-$3.7\\
{}          &            &        &  7.6 & 2.5$\pm$0.3 & 2.7$\pm$0.3 & 21 & 2.0$\times$0.8, $-$3.0\\
\hline
\textbf{200109A}\\
\txsB{}{}&  2020-02-29  & RG011  &   4.9 &   307$\pm$31 & 332$\pm$33 & 68 & 3.6$\times$2.6, 7.8\\
{}       &     2020-02-04  & BG263  & 8.4  &    380$\pm$38 & 409$\pm$41 & 106 & 2.2$\times$1.0, $-$6.3\\
{}          &             &         & 23.5 &    360$\pm$36 & 392$\pm$39 & 118 & 0.8$\times$0.3, $-$9.6\\
\hline
\textbf{201021A}\\
\rxs{}& 2020-11-05  &DD10006&  5.1 &  0.83$\pm$0.03  & 0.91$\pm$0.07  & 15 & 34.5$\times$34.5$^{*}$ \\
\pks{}& 2020-11-05  &DD10006&  5.1 & 323$\pm$16  & 334$\pm$17 & 49 & 82.4$\times$38.7, 23.9\\
\hline
\textbf{201114A} \\
\nvssB{}      & 2020-12-01/02& EG108 &  4.9  & 8.4$\pm$0.8  & 12.9$\pm$1.3 &  33  & 1.8$\times$1.1, 82.4\\
{}                 & 2020-12-06  & BG264A & 8.4  &  9.4$\pm$0.9  & 14.7$\pm$1.6 & 32 & 2.0$\times$1.0, 2.2\\
{}                 &             &        & 23.5 &  8.5$\pm$0.9  & 12.5$\pm$1.5 &  79   & 1.0$\times$0.4, $-$13.5\\
\hline
   \noalign{\smallskip}
 \end{tabular}
\begin{tablenotes}[para,flushleft]
\item  \textit{Notes:}
Col. 1 -  Candidate neutrino counterpart; 
Col. 2 -  Date of observation;
Col. 3 -  Project Code;
Col. 4 -  Observation frequency in GHz; 
Col. 5 -  Peak brightness in $\mathrm{mJy\,beam^{-1}}$;
Col. 6 -  Integrated flux density in mJy;
Col. 7 -  1-$\sigma$  noise level of the image in $\mathrm{\mu Jy\,beam^{-1}}$; 
Col. 8 -  Major axis (mas), minor axis (mas), and P.A.  (in degrees, measured from North to East) of the restoring beam.
The parameters are referred to  natural weighting images. $^{*}$ this image was produced using a circular restoring beam.
\end{tablenotes}
	\label{tab:obs_res}
\end{threeparttable}
\end{table*}

\begin{table*}[htb]
\centering
\small
\begin{threeparttable}
\caption{Spectral index measured with VLBI data.}
 \begin{tabular}{c c c c c c c c }
 \hline \hline
IC event & Source &  Date        & $\nu$            & $S_{\mathrm{peak}}$ & $uv$-range & Beam &  $\alpha$  \\ 
         &        &              & $\mathrm{(GHz)}$ & $\mathrm{(mJy\,beam^{-1})}$& M$\lambda$ & {mas$\times$mas,\deg}& \\
    \small(1) & \small(2) &\small (3) &\small (4) &\small (5) &\small (6) & \small(7) &\small (8)\\
\hline  
\textbf{190704A} & \whsp{} & 2019-07-18/22 & 1.5  & 3.1$\pm$0.3 & 2-40 & 5.6$\times$4.9, 69.9 & 0.2$\pm$0.1\\
                 &          &              &  4.4 & 3.7$\pm$0.4 &  &   & \\
\hline
                 &          &              &  4.4  & 3.4$\pm$0.4 & 5-105 & 2.8$\times$2.1 -57.8 & 0.2$\pm$0.3\\
                 &          &              &  7.6  & 3.8$\pm$0.4 &       &                       &  \\
\hline
                 &          & 2020-01-17  &  4.7   & 2.3$\pm$0.2 & 5-140  & 2.2$\times$1.1, $-$2.0 &  0.2$\pm$0.3\\
                 &          &             & 7.6    & 2.5$\pm$0.3 &        &                           & \\
\hline 
\textbf{200109A} & \txsB{} & 2020-02-29  &  4.9 & 302$\pm$30 & 4-180 & 2.2$\times$1.1, $-$6.3 & 0.4$\pm$0.3 \\ 
                 &         & 2020-02-04  &  8.4 & 378$\pm$38 & & &   \\
\hline
                 &         &   &  8.4  &374$\pm$37 & 13-250 & 1.0$\times$0.7, $-$10.4 &  $-$0.05$\pm$0.15 \\ 
                 &             &         &  23.5          & 355 $\pm$36 &  & &  \\
\hline 
 \textbf{201114A} & \nvssB{} & 2020-12-01/02& 4.9 & 7.7$\pm$0.8  & 4-165 & 2.0$\times$1.0, 2.2 &  0.4$\pm$0.3\\
                 &      &  2020-12-06      & 8.4 & 9.4$\pm$0.9 &  &  & \\
\hline
                 &      &                  & 8.4 & 9.1$\pm$0.9  & 12-250& 1.5$\times$0.9, $-$7.43 &  $-$0.02$\pm$0.14\\
                 &      &                  & 23.5 & 8.9$\pm$0.9 &  &  & \\
                 \hline
 \end{tabular}
\begin{tablenotes}[para,flushleft]
\item  \textit{Notes:}
Col. 1 -  IceCube event name; 
Col. 2 -  Candidate neutrino counterpart;
Col. 3 -  Date of  observation;
Col. 4 -  Frequency in GHz; 
Col. 5 -  Peak intensity in $\mathrm{mJy\,beam^{-1}}$;
Col. 7 -  Selected $uv$-range (in M$\lambda$) for the spectral index computation;
Col. 8 -   Major axis (mas), minor axis (mas), and P.A.  (in degrees, measured from North to East) of the restoring beam;
Col. 9 -  Spectral index.
\end{tablenotes}
	\label{tab:ind-spec}
\end{threeparttable}
\end{table*}

\begin{table*}[htb]
\centering
\begin{threeparttable}
\caption{Radio luminosity of the sources.}
 \begin{tabular}{c c c c c c c c c }
  \hline   \hline
IC event & Source & $z$   & d$_L$           & $\nu$            & $\alpha_{\mathrm{NVSS}}^{\mathrm{VLASS}}$        &$L_{\nu}$  & $\nu L_{\nu}$\\
         &        &       & $\mathrm{(Mpc)}$& $\mathrm{(GHz)}$ &         &$\mathrm{(W Hz^{-1})}$ & $\mathrm{(erg\, s^{-1})}$\\
\small(1) & \small(2) &\small (3) &\small (4) &\small (5) &\small (6) & \small(7) &\small (8)\\
\hline
 190704A & \whsp{} & 1.914 &  14722.8 & 1.5 &  $-$0.2$\pm$0.3   & (5.4$\pm$1.0)$\times10^{25}$ &  (8.0$\pm$1.7)$\times10^{41}$ \\
                &             &       &          &  4.4 &                & (4.2$\pm$1.1)$\times10^{25}$ &  (2.3$\pm$0.5)$\times10^{42}$\\
\hline
                &             &       &          &  4.7 &                & (3.8$\pm$0.7)$\times10^{25}$ &  (1.7$\pm$1.2)$\times10^{42}$\\
\hline  
 200109A & \txsB{}&  0.91  &  5880.5 & 4.9 &  0.08$\pm$0.19  & (6.4$\pm$1.2)$\times10^{26}$ & (2.8$\pm$0.5)$\times10^{43}$ \\
\hline  
 201021A & \pks{} &  0.586  &  3430.4   & 5.1 & 0.06$\pm$0.19  &  (2.7$\pm$0.3)$\times10^{26}$ & (1.2$\pm$0.1)$\times10^{43}$ \\
 \hline
 \end{tabular}
 \begin{tablenotes}[para,flushleft]
\item \textit{Notes:} Col. 1 -  IceCube event name; 
Col. 2 -  Candidate neutrino counterpart;
Col. 3 -  Redshift of the source;
Col. 4 -  Luminosity distance in Mpc; 
Col. 5 -  Frequency in GHz; 
Col. 6 -  Spectral index measured between  1.4 GHz and 3 GHz (from NVSS and VLASS data);
Radio luminosity resulted from our observations, expressed   in $\mathrm{W Hz^{-1}}$ (Col. 7) and in $\mathrm{erg s^{-1}}$  (Col. 8).
 \end{tablenotes}
\label{tab:lum}
\end{threeparttable}
\end{table*}

\subsection*{Notes on individual neutrino events}

\paragraph{\textbf{\icnA{}}}

 There are two possible  $\gamma$-ray      counterparts,  \gammajtfn{} and  \gammajtff{},  spatially coincident with \icnA{}.  The former is a    4FGL  $\gamma$-ray source without any associated low-energy counterpart (Table~\ref{tab:gammaray}).  
The only radio source within its error ellipse is  \nvssA{}. Taking into account the unremarkable radio and optical properties of this source,  we only observe it with a few scans.  
 The resulting   characteristics of this source are described in Appendix~\ref{appendix}.

The other $\gamma$-ray source, \gammajtff{}, was a new detection at the time of the follow-up campaign \citep{2019ATel12906....1G}, later included in the 4FGL-DR2 catalog (Table~\ref{tab:gammaray}). The possible  counterpart   is \whsp{}, at redshift 1.914, classified as HBL \citep{2015A&A...579A..34A,2017A&A...598A..17C}. 
A possible  connection between  HBL sources and archival IceCube neutrino events have been suggested by   \citet{2020MNRAS.497..865G} and \citet{2022MNRAS.510.2671P}. 

\begin{figure}
    \centering
    \includegraphics[height=0.45\textwidth]{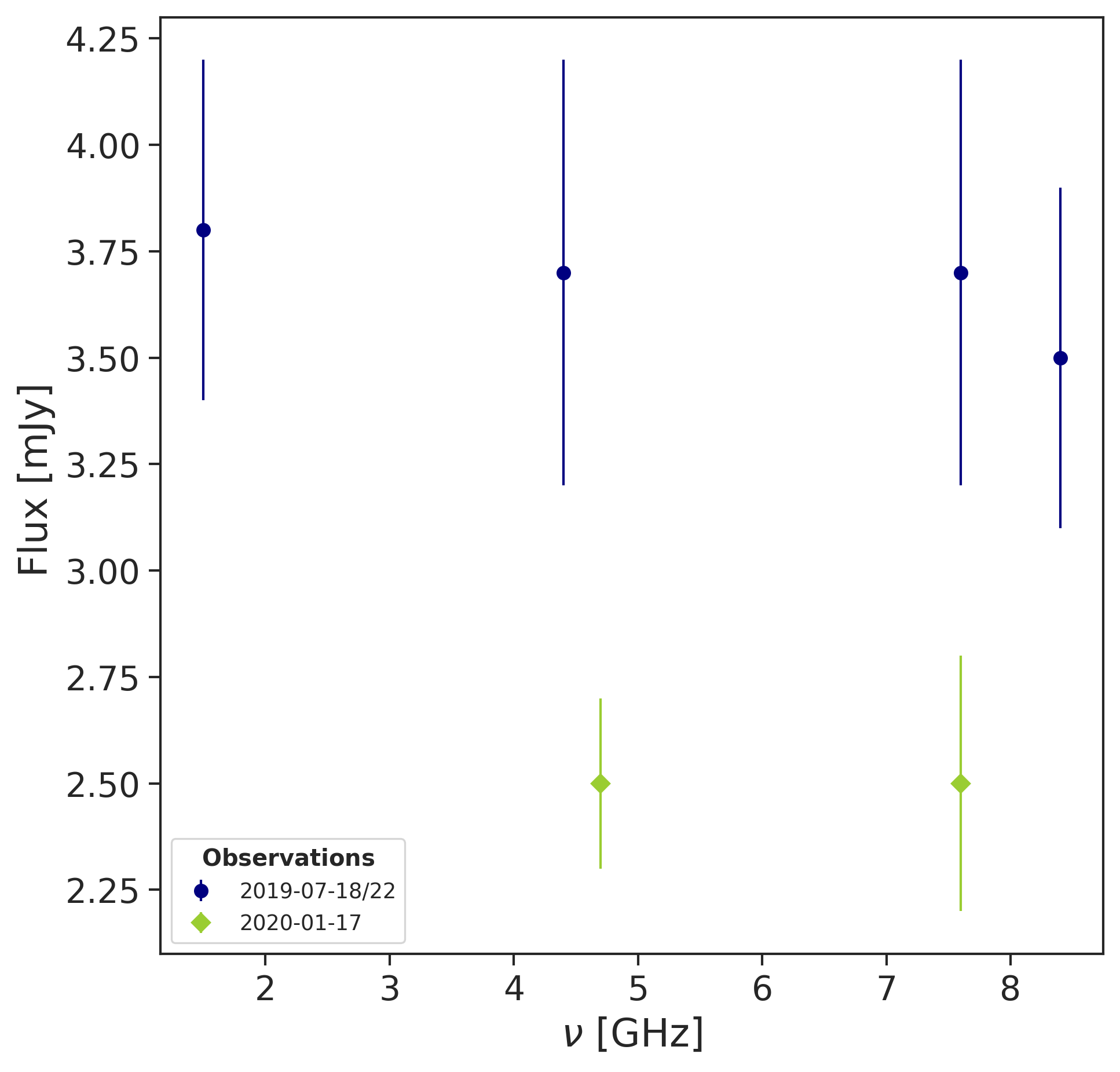}
\caption{   \whsp{} total radio flux density at different frequencies and epochs.
    \label{fig:1045sp_ind}
}
\end{figure}

\whsp{} is unambiguously detected as a compact single component at all the frequency bands   in the  two epochs. 
Details of the radio images  are listed in Table~\ref{tab:obs_res} and the total flux density at each frequency  in the two epochs is shown in Fig.~\ref{fig:1045sp_ind}.  
To determine if the source  underwent an increase of flux, we  compared  the VLBA integrated flux densities in the two epochs, measured at around 4 GHz and at 7.6 GHz, using  Eq.~\ref{eq:var}. The resulting  variability index $V \sim$ 0.03 between the epochs at both the frequencies implies that the  source did not experience an increased radio activity  after the neutrino detection.
 It was not possible  to check  the source long-term activity level because there are no archival RFC data   available. The  peak intensity and the total flux density are consistent with each other in both our observations, indicating   that \whsp{} is a compact source at   VLBA-scales. By fitting the emission observed in the   second epoch with an elliptical component using the \texttt{imfit} task, the major axis  (deconvolved from the beam) turns out to be  about 6 pc at 4.7 GHz and 7.6 GHz. In the first epoch, the dimensions of the source were not retrieved because it appears as point-like when deconvolved from the beam. From the 4.4 GHz \texttt{imfit} analysis we obtained an upper limit for the major axis  $\le$ 14 pc. 

The surveys data at 1.4 GHz  indicate that the source remains compact on arcsecond-scales  (Table~\ref{tab:nvss-first}).
The comparison of arcsecond and mas-scales (even if not simultaneous) does not suggest the presence of any emission on intermediate scales. Carried out a month before the neutrino event, the VLASS observation  recorded flux densities in agreement with the NVSS measurements, slightly lower  than FIRST ones though consistent within the uncertainties. 

From our  multi-frequency data, we could compute the synchrotron spectral indices of \whsp{} which is $\sim$0.2 both, between 1.5 GHz and 4.4 GHz and between 4.4 GHz and 7.6 GHz (Table~\ref{tab:ind-spec}). The  compact structure of the source remains self-absorbed, at least until 7.6 GHz, in both the epochs. Accounting for the source redshift,   the radio luminosity     is  $\sim$4.2 $\times10^{25}$W/Hz at 4.4 GHz  in the first epoch and $\sim$3.8$\times10^{25}$W/Hz at 4.7 GHz in the second one (Table~\ref{tab:lum}). 



\paragraph{\textbf{\icnB{}}}\label{icnb}

\begin{figure}
    \centering
    \includegraphics[height=0.45\textwidth]{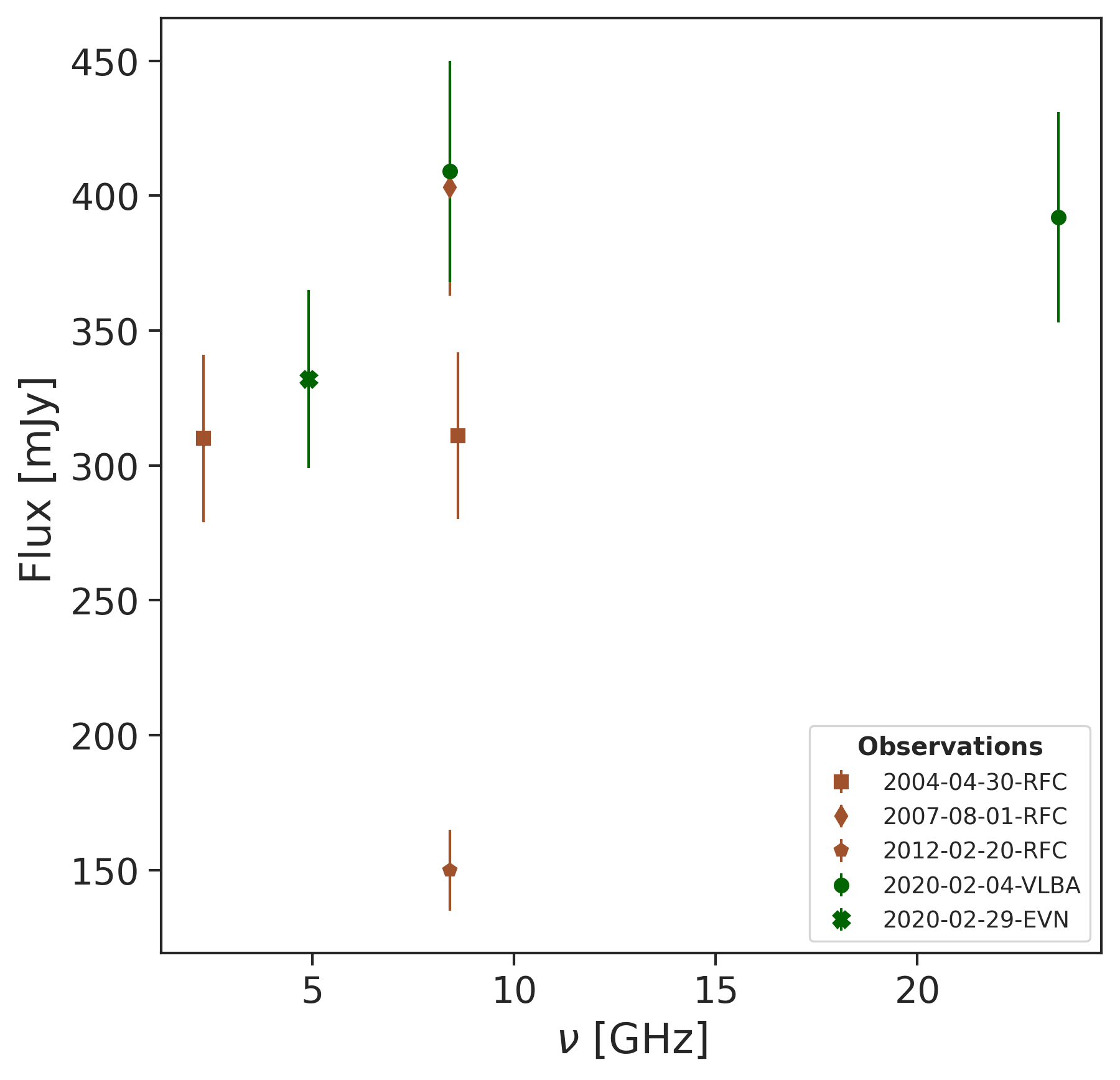}
\caption{   \txsB{} total radio flux density at different frequencies and epochs.
    \label{fig:1100sp_ind}
}
\end{figure}

\begin{figure}
    \centering
    \includegraphics[height=0.4\textwidth]{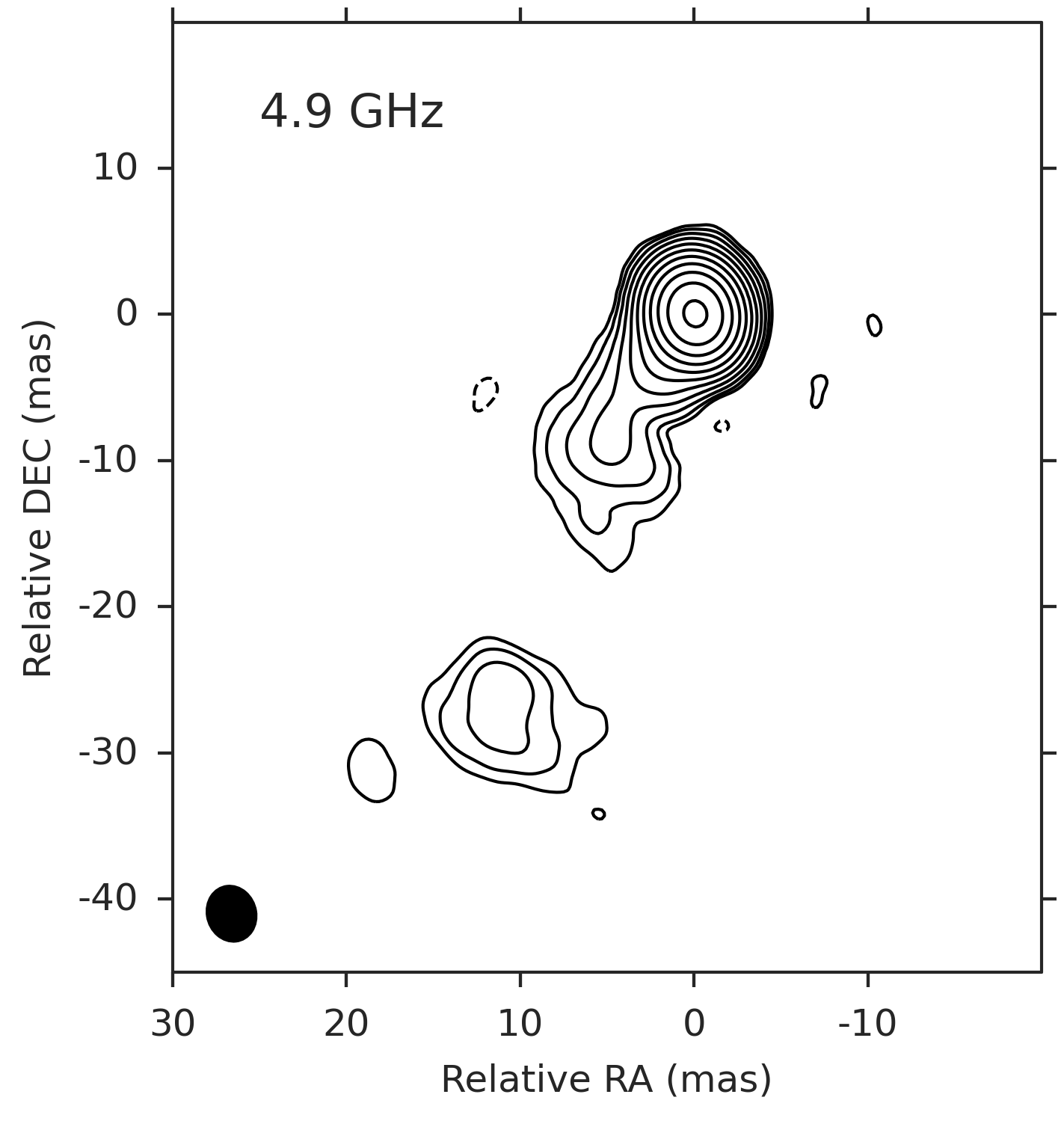}
    \includegraphics[height=0.4\textwidth]{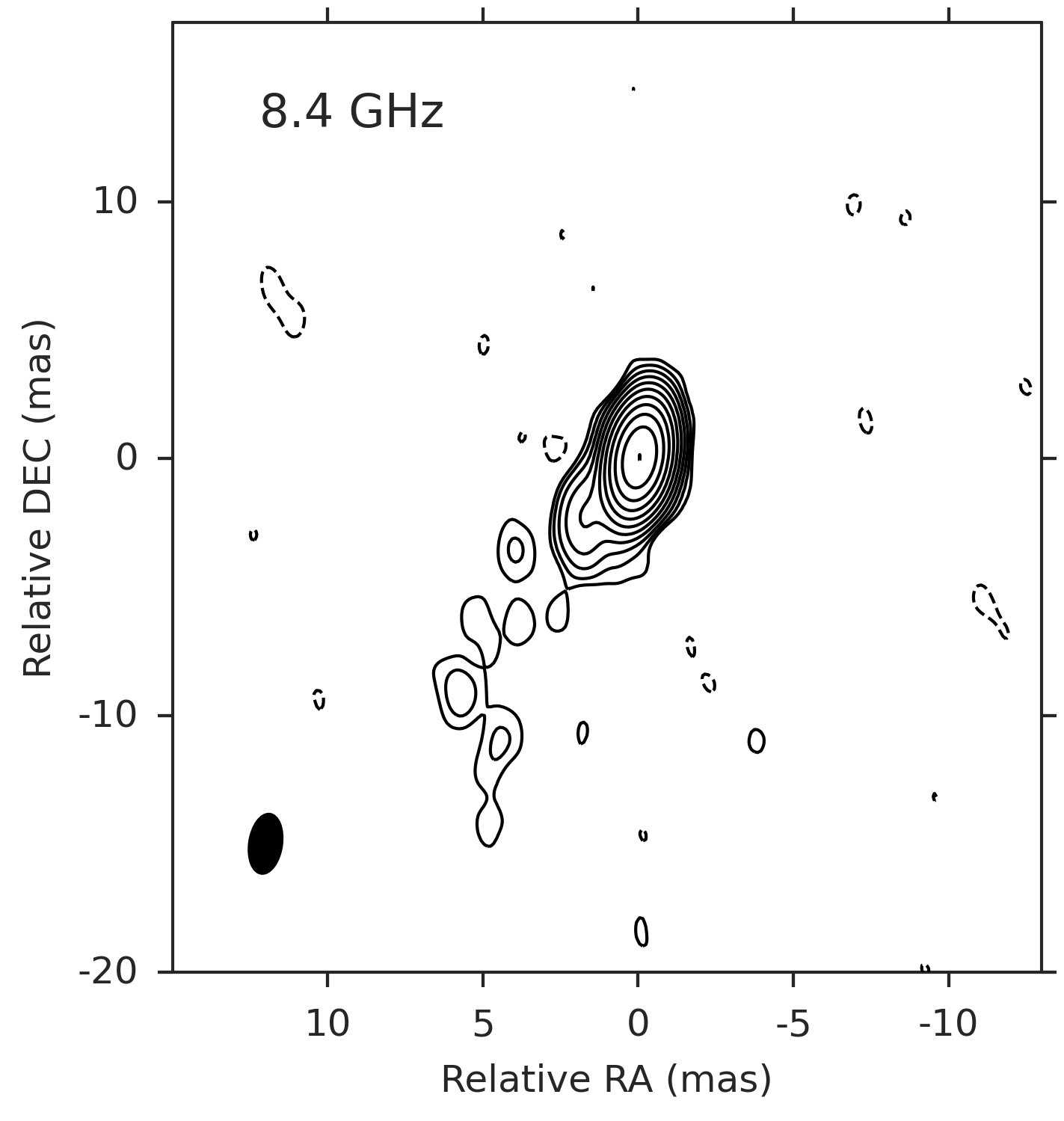}\,
    \includegraphics[height=0.41\textwidth]{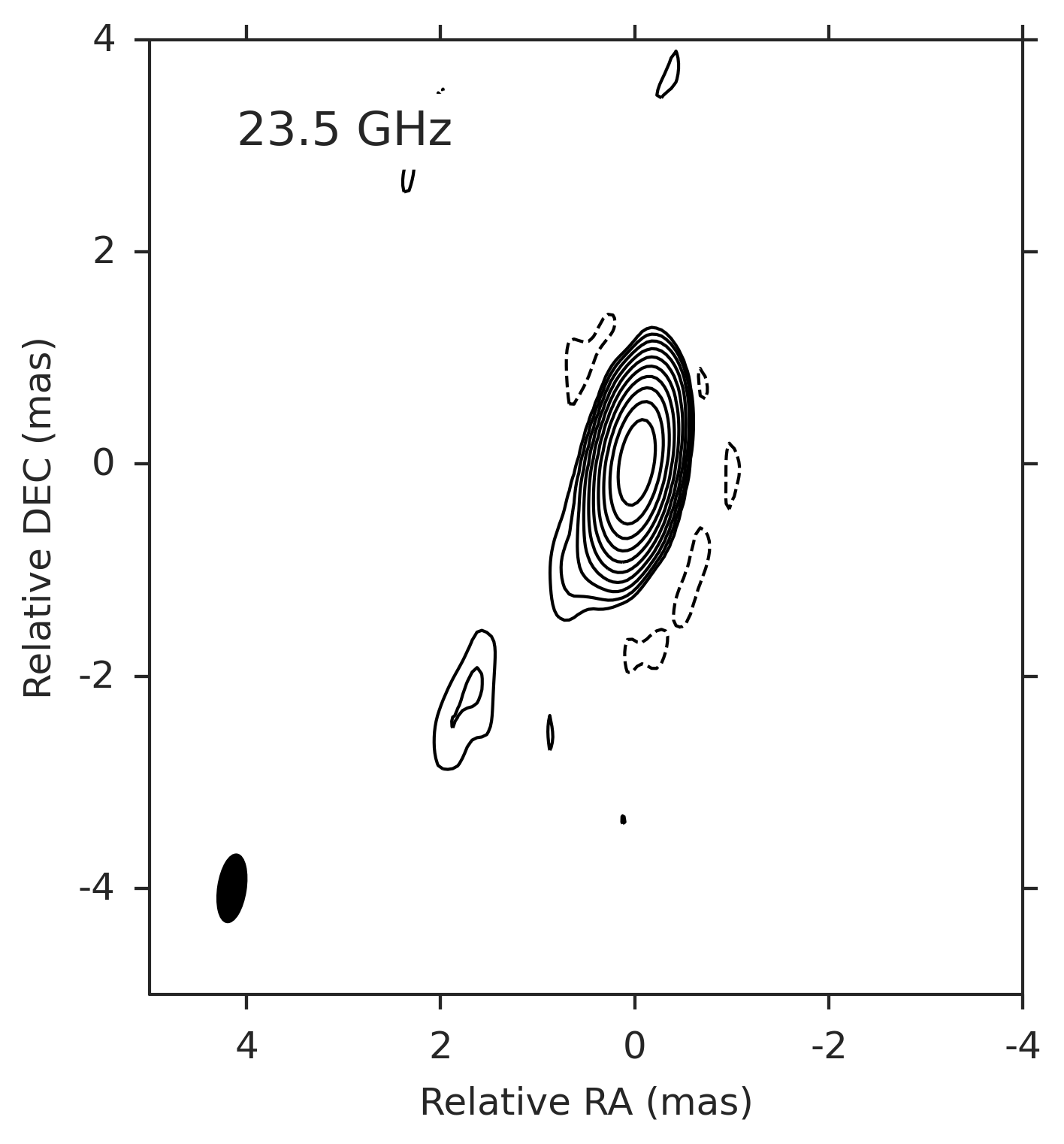}
\caption{EVN-4.9 GHz and VLBA-8.4 GHz and VLBA-23.5 GHz  contour images of \textbf{\txsB{}}. The contour levels are drawn from 3 $\times$ the rms noise of the images. Contours increase by a factor of 2.
The noise level and the beam size of each image are reported in columns 7 and 8 of Table~\ref{tab:obs_res}.
The black ellipse in the bottom-left corner represents the restoring beam.
    \label{fig:1100+122}
}
\end{figure}

Among the three candidate $\gamma$-ray counterparts of  the Gold event \icnB{},  the 4FGL catalog reports   high statistically significant association only for one of them, J11103.0+1157, associated with  the   FSRQ \txsB{}. The other two  $\gamma$-ray  candidates, J1114.6+1225 and J1055.8+1034, can be spatially associated with  \icnBone{} and \icnBtwo{}, respectively. However, these associations are not confirmed by statistical arguments as the Likelihood Ratio method adopted for source associations in the 4FGL \citep{2020ApJS..247...33A}. 
 \icnBone{} is located at about 4 degrees from the neutrino best-fit position, making its connection with the event less likely compared to the other two candidates.  Moreover,  both  \icnBone{} and \icnBtwo{} are not classified as blazar or blazar candidates, thus we present  the VLBI analysis of those two targets  in Appendix~\ref{appendix}, while  in this Section we focus on \txsB{} which was immediately pointed out as  potential neutrino source by \citet{2020ATel13397....1K}. 
The final  images for the VLBA and EVN observations of this source are shown in Fig.~\ref{fig:1100+122}.   The images  parameters are listed in Table~\ref{tab:obs_res}.  The total flux density  measured from our data with respect to the  total flux density from RFC data  can be seen in Fig.~\ref{fig:1100sp_ind}.

An elongated structure extending towards the south-east direction is recognizable  at all the observing  frequencies. 
We modeled the source structure  with Gaussian components with the modelfit procedure in DIFMAP for the new and archival VLBI observations.  The properties of the components are reported in  Table~\ref{tab:mf_txs} in the Appendix~\ref{modelfit}.  At  each  frequency we fitted an elliptical/circular component representing the compact core emission.  In addition to the core, five, six, and three   other Gaussian models describe the 4.9 GHz, 8.4 GHz, and 23.5 GHz jet structure, respectively.    The maximum elongation measured  as the distance between the center of the core component and the center of the outermost component  is about 255 pc, 150 pc and 22 pc, at 4.9 GHz, 8.4 GHz, and 23.5 GHz  respectively (7.8 pc/mas).   The properties of the RFC observations are summarized in Table~\ref{tab:rfc}. 
 The  best-fit parameters of the components  are reported  in Table~\ref{tab:mf_txs}. The    model components identified in our observation are not clearly cross-identified in the RFC data. This can be due to either more than 10 years elapsed between the   observations or the absence of intrinsically distinct compact regions in the jet or the different data quality. 
The absence of well defined, compact components or stationary components (i.e. found at the same radius in different epochs) indicates that we are sampling a smooth, featureless and quite homogeneous  jet emission. 
 The jet lies at a P.A. between  $\sim$ 140 and 165 degrees (measured from North to East with respect to the image central pixel) and it does not show any   bending. Only the last  component  identified in our 8.4 GHz data  seems to be misaligned with respect to the others, detected at a P.A. of 170 degrees. It might  represent a  curved structure  undetected in the other observations.  
 
In Table~\ref{tab:ind-spec} we report the spectral index  of \txsB{} measured from VLBI data. The source shows an inverted spectrum  with a  peak around 8.4 GHz. In the 8.4$-$23.5 GHz frequency range the spectrum is flat.  A flatter spectrum  at  lower frequencies (between 2.3 GHz and 8.4 GHz) results   from the archival RFC data (Table~\ref{tab:rfc_ind-spec}).   
Our spectral index measurements disagree with the spectral behaviour observed with the RATAN-600 telescope, reported by \citet{2020ATel13405....1K}.  The synchrotron spectrum derived from the RATAN-600 observation is inverted up to 22 GHz. In fact,   they measured the highest flux density,  552$\pm$39 mJy, at 22 GHz. Around this frequency we observed a flux density of $\sim$390 mJy.  RATAN-600 observation was performed  on 11 and 14 January 2020, only few days after the neutrino detection, while our follow-up was carried out more than one month after the detection. The  different  flux density measurements around 22 GHz and 23.5 GHz  could suggest that the radio-flare from \txsB{} occurred before or simultaneously to the neutrino emission and we observed the source when the flare was already extinguished. As reported in the NRAO notice mentioned at the beginning of this Section,  because the BG263 observation is carried out in DDC mode, the flux density of \txsB{} at 23.5 GHz could be   12\% lower than the one that we have measured. By adding a factor  12\% to   Eq.~\ref{eq:errorfluxradio} (squaring sum), we obtain a flux density of  392$\pm$61, so the upper limit would be 453 mJy. The 22 GHz RATAN-600 lower limit is 513 mJy, which results in a non-negligible difference of about 60  mJy between our and RATAN-600 estimation. 
  Part of this discrepancy may arise from the difference in spatial scales to which RATAN-600 and VLBA are sensitive. Indeed, diffuse emission contributes to the total flux density measured by RATAN-600 while it is filtered out by VLBA. For this reason, the flux densities measured by the two instruments must always obey the relation $S_\mathrm{RATAN-600} \ge S_\mathrm{VLBA}$. However, it is not possible to determine if the discrepancy of 60 mJy is entirely attributable to this effect or if it indicates an actual decrease of the flux density of the source in the time range between the RATAN-600 and VLBA observation.


We evaluated the variability of the source with respect to the RFC data taken at 8.6 GHz, which is the frequency closest to our observation frequency.  We obtained $V=0.38$ between 2012 and 2020 (Table~\ref{tab:variab}).  
 From NVSS and FIRST data,    \txsB{}   flux densities at 1.4 GHz are consistent to each other and quite lower than our observations  (Table~\ref{tab:nvss-first} and Table~\ref{tab:obs_res}). Although the different  resolution and observing frequency  do not allow for a rigorous comparison between the NVSS/FIRST   results and the  VLBI measurements, a simple explanation for the  discrepancy in the flux density is offered by  the   inverted  spectrum in the low frequencies regime (see Table~\ref{tab:ind-spec}), which implies low flux density at that frequencies. Adopting the spectral index measured from our data, the flux density extrapolated at 1.4 GHz results in agreement with the one inferred from   the surveys data. The presence of emission on intermediate angular scales seems unlikely. 
Carried out  two years prior  the neutrino detection, the first epoch   VLASS  observation of \txsB{} reveals  flux density  measurement in agreement with the  NVSS and FIRST results.  In  the second VLASS epoch (few months after the neutrino detection and after our  observations) the source shows higher flux density compared to the first VLASS epoch. The two measurements are however consistent with each other within the uncertainties  ($V=-0.05$). This   suggests there is no significant   variability on arcsecond-scales, but we also remark that the VLASS data are still preliminary and should not be over-interpreted.

\begin{table*}[htb]
\centering
\small
\begin{threeparttable}
\caption{RFC Spectral index.}
 \begin{tabular}{c c c c c c c c}
 \hline \hline
IC event & Source & Date       & $\nu$           &   $S_{\mathrm{peak}}$               & $uv$-range & beam  & $\alpha$\\ 
{}       &        &            &$\mathrm{(GHz)}$ & $\mathrm{(mJy\,beam^{-1})}$&   M$\lambda$& {mas$\times$mas,\deg}&\\
\small(1) & \small(2) &\small (3) &\small (4) &\small (5) &\small (6) & \small(7) &\small (8)\\
\hline 
 200109A & \txsB{} & 2004-04-30 & 2.3 & 254$\pm$26  & 6-70 & 3.1$\times$2.5, 0.5 & 0.1$\pm$0.1  \\ 
         &         &            & 8.4 & 302$\pm$30 &   &      &                         \\
201114A  & \nvssB{}& 2013-04-08/09  & 4.3  & 18$\pm$2 & 4-110 & 2.9$\times$1.6, 5.5  &  0.05$\pm$0.25 \\
         &            &  & 7.6  & 19$\pm$2 &  & & \\
         \hline
 \end{tabular}
\begin{tablenotes}[para,flushleft]
\item \textit{Notes:}
Col. 1 -  IceCube event name; 
Col. 2 -  Candidate neutrino counterpart;
Col. 3 -  Date of observation;
Col. 4 -  Frequency in GHz; 
Col. 5 -  Peak intensity in $\mathrm{mJy\,beam^{-1}}$;
Col. 6 -  Selected $uv$-range (in M$\lambda$) for the spectral index computation;
Col. 7 -  Major axis (mas), minor axis (mas), and P.A.  (in degrees, measured from North to East) of the restoring beam;
Col. 8 -  Spectral index.
\end{tablenotes}
	\label{tab:rfc_ind-spec}
\end{threeparttable}
\end{table*}

\begin{table*}[htb]
\centering
\small
\begin{threeparttable}
\caption{Variability index calculated between  our VLBI observations and RFC observations.}
 \begin{tabular}{l c c c c c c c c c }
 \hline \hline
Source & Obs. & Date       & $\nu$            &   beam  &  {$S_{\mathrm{int}}$}       &  V \\ 
{}     &    &   & $\mathrm{(GHz)}$ & {mas$\times$mas,\deg} & {$\mathrm{(mJy)}$} &\\
\small(1) & \small(2) &\small (3) &\small (4) &\small (5) &\small (6) & \small(7) \\
\hline 
\textbf{200109A}\\
TXS 1100+122& RFC  & 2004-04-30 & 8.6 & 2.0$\times$0.9, 1.1 & 311$\pm$31 &  \\
            & VLBA & 2020-02-04  & 8.4 &  2.2$\times$1.0, $-$6.3 & 409$\pm$41 & 0.04 \\
            \hline
            & RFC   & 2007-08-01 & 8.6  & 2.2$\times$1.2, 35.1 & 403$\pm$41 &   \\
            & VLBA  & 2020-02-04 & 8.4 &  2.2$\times$1.0, $-$6.3 & 409$\pm$41 & $-$0.09 \\
\hline
            & RFC   & 2012-02-20 & 8.6  & 2.0$\times$0.9, 8.0 & 150$\pm$15&   \\
            & VLBA  & 2020-02-04 & 8.4 &  2.2$\times$1.0, $-$6.3 & 409$\pm$41 & 0.38  \\
\hline
\textbf{201114A} \\
\nvssB{} & RFC       & 2013-04-08/09 & 4.3 &  4.6$\times$1.9, $-$7.3  &  22.5$\pm$2.4  & \\ 
             & EVN       & 2020-12-01/02 & 4.9 & 1.8$\times$1.1, 82.4 & 12.9$\pm$1.3  &  0.17 \\
             \hline
             & RFC       & 2013-04-08/09 & 7.6 &  2.8$\times$1.1, $-$12.2  &  23.2$\pm$2.6 &  \\ 
             & VLBA      & 2020-12-06    & 8.4 & 2.0$\times$1.0, 2.2  & 14.7$\pm$1.6 & 0.12 \\
             \hline 
             & RFC       & 2013-10-19 & 7.6 &  2.2$\times$1.3, $-$3.3   &  15.2$\pm$1.6 &  \\ 
             & VLBA      & 2020-12-06 & 8.4 &  2.0$\times$1.0, 2.2  & 14.7$\pm$1.6 &   $-$0.09 \\
                      \hline

 \end{tabular}
\begin{tablenotes}[para,flushleft]
\item \textit{Notes:}
Col. 1 -   Candidate neutrino counterpart;
Col. 2 -  Origin of the VLBI observation: RFC or our VLBA/EVN observations;
Col. 3 -  Date of observation;
Col. 4 -  Frequency in GHz; 
Col. 5 -  Major axis (mas), minor axis (mas), and P.A.  (in degrees, measured from North to East)  of the restoring beam;
Col. 6 -  Integrated flux density in mJy;
Col. 7 -  Variability index.
Negative values of $V$ means that the source does not show variability \citep{1992ApJ...399...16A}.
\end{tablenotes}
	\label{tab:variab}
\end{threeparttable}
\end{table*}

\paragraph{\textbf{\icnC{}}}

Two $\gamma$-ray sources have been detected as possible   \icnC{} counterpart.   We performed the e-MERLIN follow-up of this event and the resulting images parameters are reported in  Table~\ref{tab:obs_res}. 

\begin{figure}
    \centering
    \includegraphics[height=0.42\textwidth]{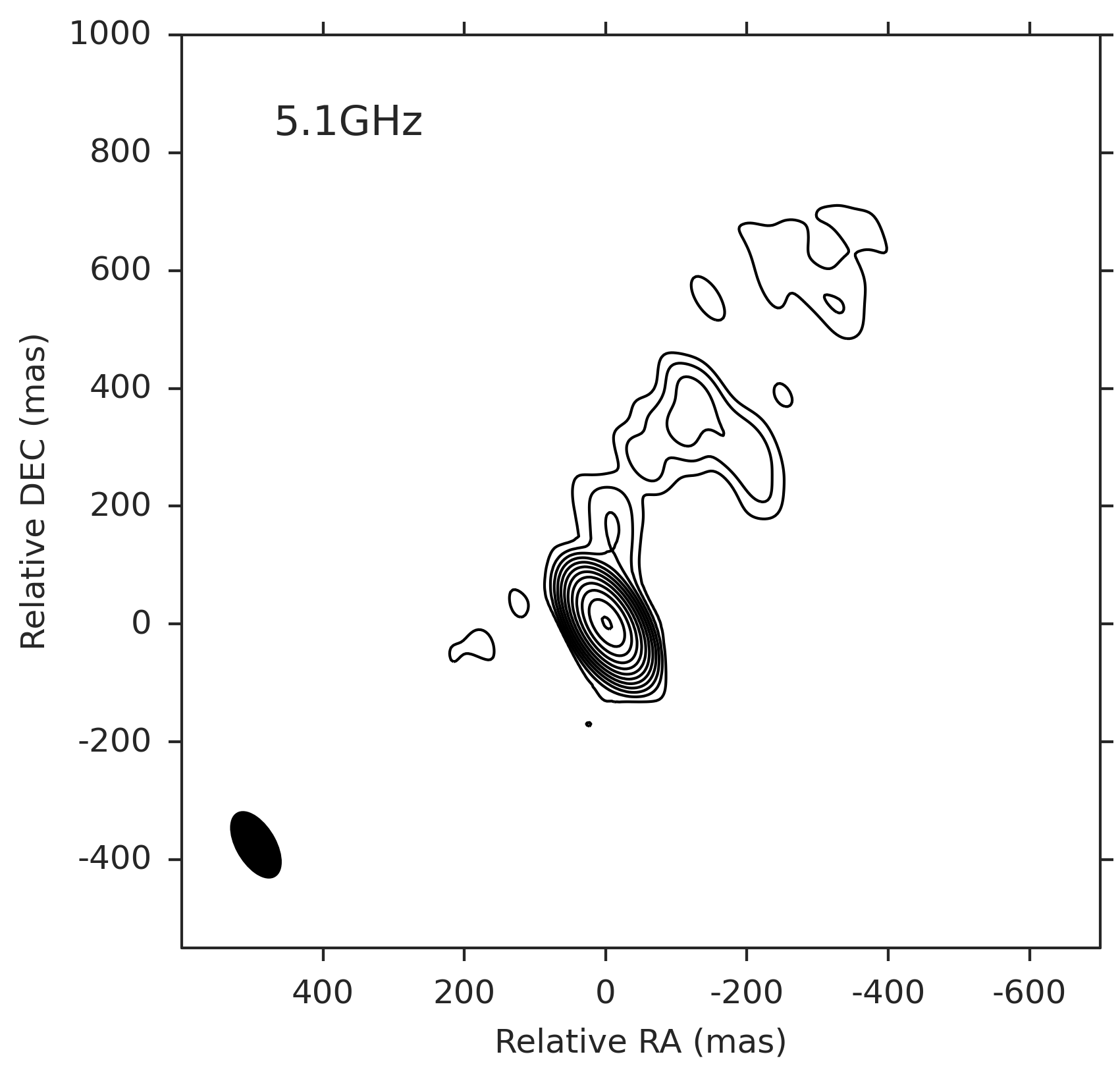}
\caption{e-MERLIN    contour image  of \textbf{\pks{}}. The contour levels are drawn from 3 $\times$ the rms noise of the images and increase by a factor of 2.
The noise level and the beam size of the image are reported in columns 7 and 8 of Table~\ref{tab:obs_res}.
The black ellipse in the bottom-left corner represents the restoring beam.
    \label{fig:1728}
}
\end{figure}

\begin{itemize}
    \item \textit{\pks{}} lies about 70 arcmin outside the 90\% localization region of the event. However, being a bright FSRQ experiencing a temporal coincident   high state  at 15.3 GHz (from the MOJAVE data, described below), it would represent a good candidate for the neutrino association. Our e-MERLIN observation  shows a jet structure pointing in the north-west direction (Fig.~\ref{fig:1728}). 
    The   emission extends for about 700 mas from the core, which corresponds to a distance of $\sim$4.6 kpc (6.6 pc/mas).  The jet remains collimated along this distance. 
    From the 5.1 GHz e-MERLIN observation, it results that the  emission from the core region dominates over the total flux density of the source. 
    The same emerges from survey  observations which sample arcsecond-scales (Table~\ref{tab:nvss-first}).  In addition, our 5.1 GHz  and the surveys 1.4 GHz and 3 GHz flux density are comparable to each other  within the errors, indicating the dominance of the core emission holds from the smaller scales over  larger ones. The modelfit analysis of the e-MERLIN visibility data indicates the jet contributes less than 5\% to the total emission while the largest contribution is given by the compact core component.    Best-fit  model parameters  are reported in Table~\ref{tab:mf_pks} in Appendix~\ref{modelfit}.   
    
    Being \pks{} a bright VLBI calibrator, it has been frequently monitored with VLBI observations over the years. We reported the last five 15.3 GHz-MOJAVE observations taken   in epochs  close  to the neutrino detection, with one precisely conducted on the day of the detection (Table~\ref{tab:rfc}). 
    The modelfit analysis that we performed on the MOJAVE data (Table~\ref{tab:mf_pks}) does not highlight    long-standing features in the jet between 2018 and 2020.   
 

    The 15.3 GHz flux density is  higher than our almost simultaneous 5.1 GHz observation  due to the inverted shape of the self-absorbed synchrotron spectrum of the core. 
    As a consequence of the different scales sampled by the e-MERLIN and VLBI data, we could not calculate the spectral index   between 5.1 GHz and 15.3 GHz in a proper way, despite the data were taken almost simultaneously. The $uv$-range of the two dataset does not overlap.
    To set an upper limit to the core  spectral index  using the available data, we first identify the core emission region in the e-MERLIN data. 
    To do this, we fit the visibility points with a Delta function and an extended elliptical component representing the jet contribution. 
    The jet component extends on angular scales which are filtered out by the VLBI observations, then the e-MERLIN Delta function alone can be considered as the core emission region. 
    The flux density of this component is 323$\pm$16 mJy. We then  interpolated the 15.3 GHz-MOJAVE   data    assuming a linear growth with time between the last MOJAVE epochs and we  retrieved the 15.3 GHz flux density value at the date of our e-MERLIN observation. This results to be  572$\pm$57 mJy. Finally, the calculated 5.1 GHz to 15.3 GHz spectral index is   0.5$\pm$0.1.
    
    The variability index of the source  at 15.3 GHz   turns out to be $\sim 0.1$ in the time range close to the neutrino event (2020-05 vs 2020-12), showing  that an increased  activity is detected at least in the nuclear region.


\item \textit{\rxs{}}  is the possible counterpart for the newly detected $\gamma$-ray source,   J1725.5+1312, reported in \citet{2020ATel14111....1B}. This  $\gamma$-ray source was significantly detected in the   $\gamma$-ray band only performing    the integration of the \fermi{} data taken  over the period of 10 years \citep{2020ATel14111....1B}. This suggests the idea that   J1725.5+1312 is a faint source in the  $\gamma$-ray band, experiencing high and low level of activity. Also, the association with a  radio (although  weak)    counterpart,   identified in the VLASS data, indicates  it represents a good blazar candidate.

\rxs{} is clearly detected in our e-MERLIN images. Its radio structure consists of a main component with a flux density of about 0.96 mJy, and a blob component 90 mas south-east from the core, with a flux density of about 160 $\mu$Jy. The source is not detected in the NVSS, allowing us to set an upper limit of about 0.6 mJy (i.e., 3 times the noise measured on the image plane).  No FIRST data are available for \rxs{}, being the source slightly outside the region covered by the survey. In the VLASS image \rxs{}   appears as a compact source with a flux of 0.9$\pm$0.1 mJy  (Table~\ref{tab:nvss-first}), consistently with our result. 
No  indication about the \rxs{} variability can be deduced from the available data. 


\end{itemize}

\paragraph{\textbf{\icnD{}}}
 
 \begin{figure*}[htb]
    \centering
    \includegraphics[height=0.3\textwidth]{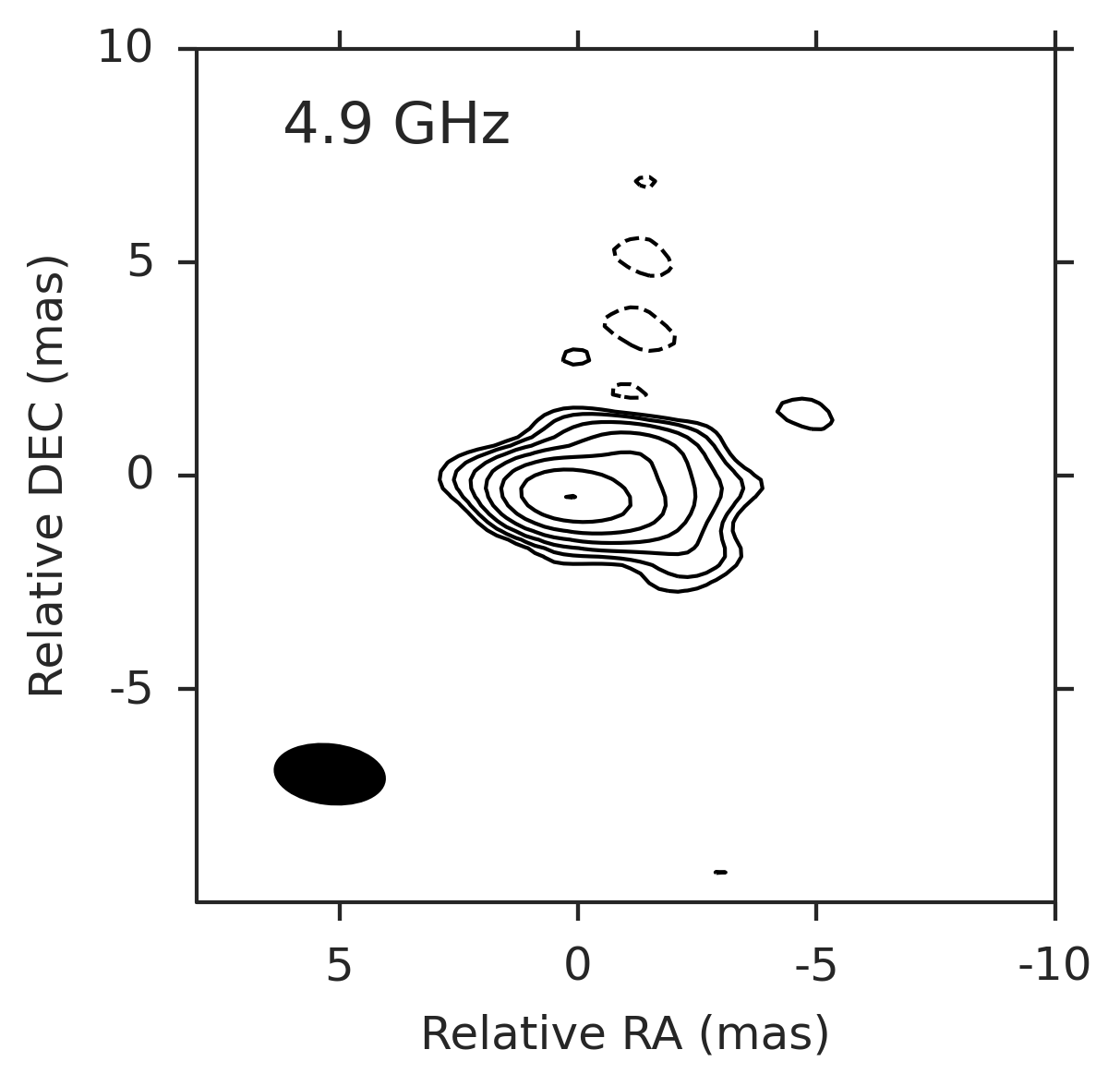}
        \includegraphics[height=0.3\textwidth]{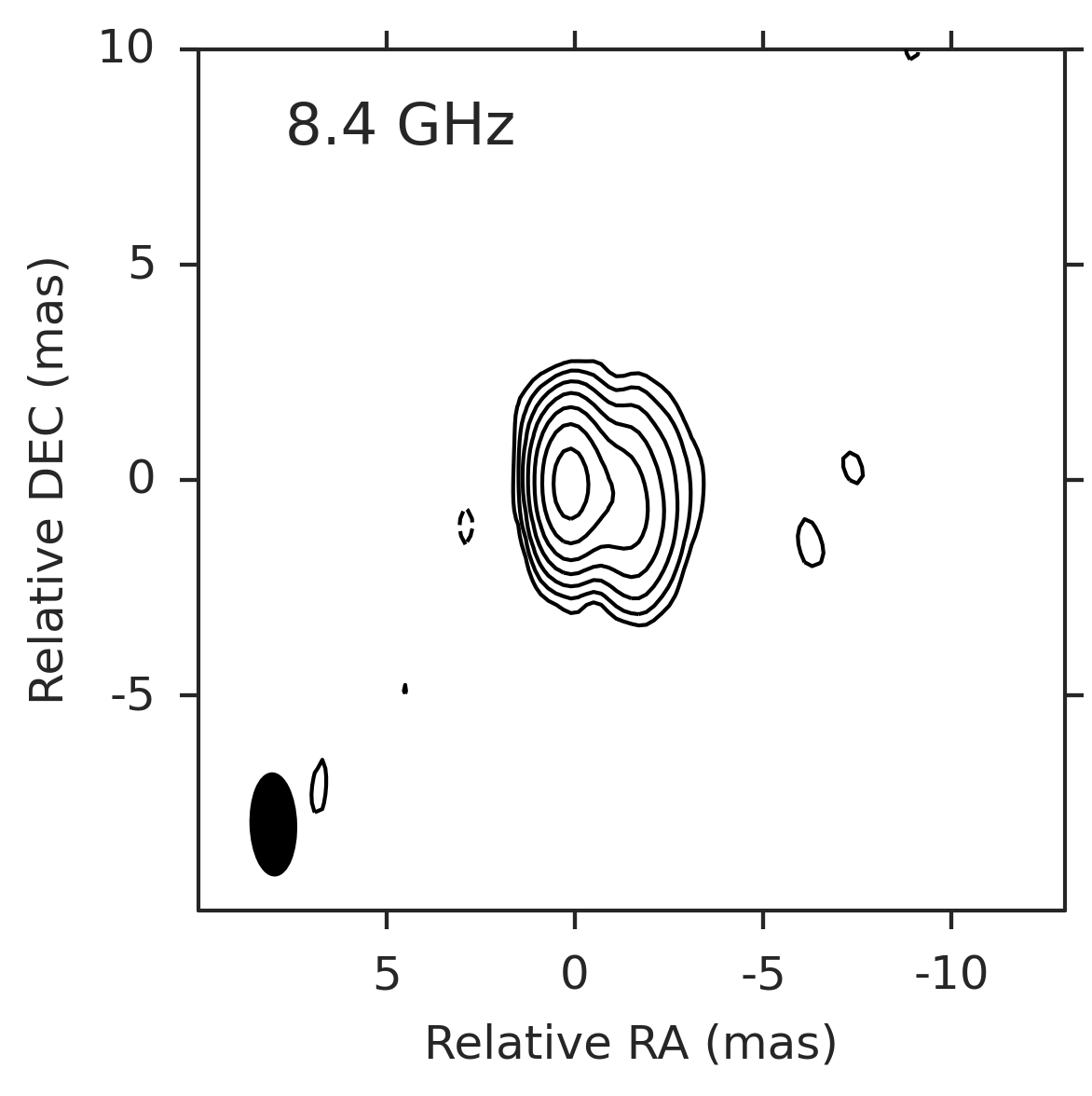}
    \includegraphics[height=0.295\textwidth]{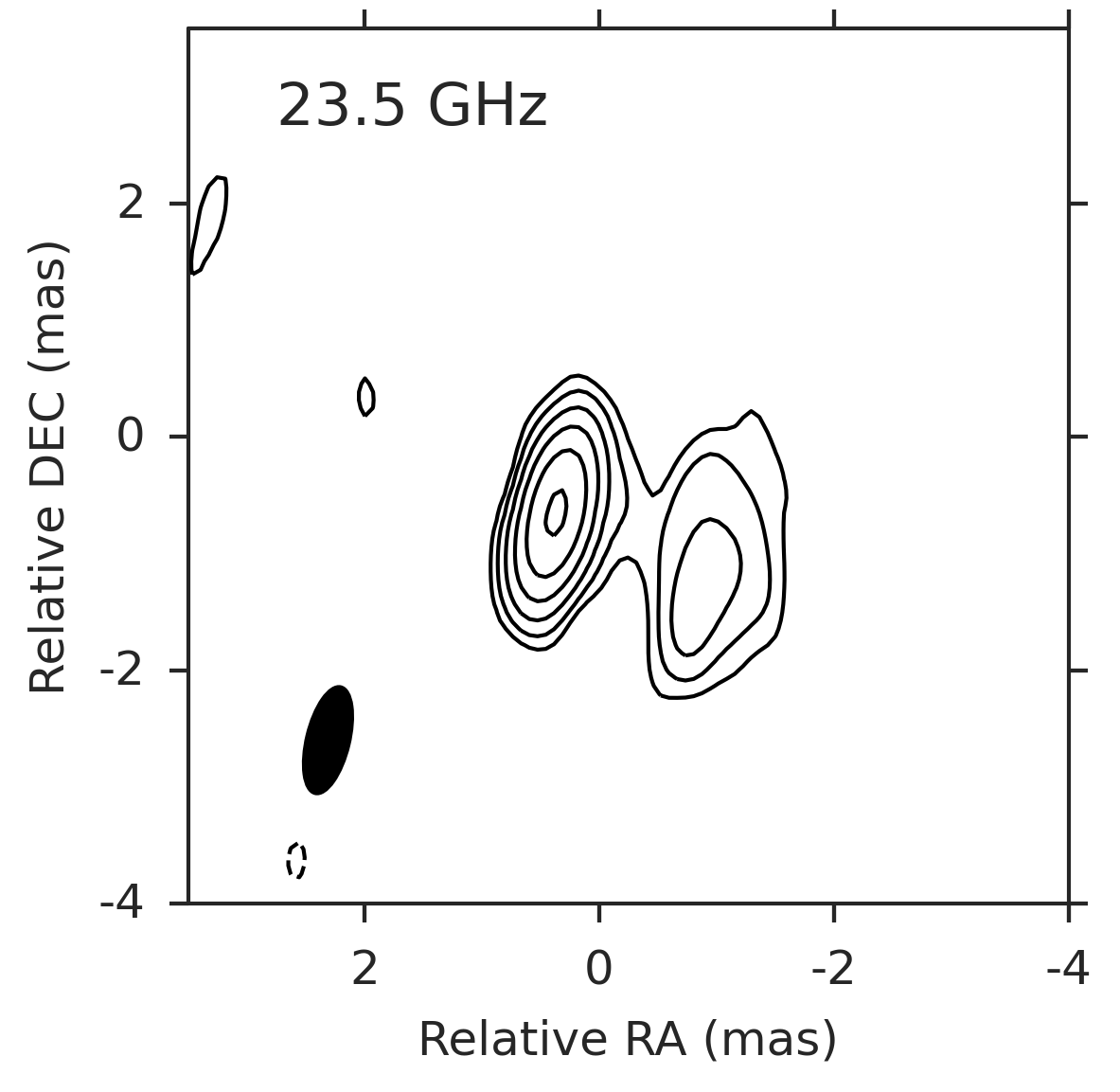}
    \caption{EVN-4.9 GHz and VLBA-8.4 and -23.5 GHz   contour images of \textbf{\nvssB{}}. The contour levels are drawn starting from 3  $\times$ the rms and are spaced by power of two.
The noise level and the beam size of each image are reported in columns 7 and 8 of Table~\ref{tab:obs_res}.
The black ellipse in the bottom-left corner represents the restoring beam.
    \label{fig:0658}
}
\end{figure*}

The \icnD{} Gold event was found in spatial coincidence with three  possible counterparts, two of them are known blazars  without an associated $\gamma$-ray source (Table~\ref{tab:counterpart}).  Here we focused on the only candidate with a  $\gamma$-ray association: \nvssB{}, which  has been also targeted  by a  multi-wavelength campaign started  after  the neutrino detection \citep[de Menezes et al. in prep]{deMenezes:2021Fl}. A low-significance excess of archival low-energy neutrinos observed by IceCube and spatially consistent with the source is reported in \citet{2019JCAP...02..012H}. Moreover, a 155 GeV photon from the $\gamma$-ray counterpart of \nvssB{}  has been
detected by the \fermi-LAT on 2018-01-28 \citep{2020ATel14200....1B}. The
$\gamma$-ray source is also included in the Third \fermi-LAT Catalog of High-Energy Sources \citep[3FHL;][]{2017ApJS..232...18A}, suggesting it as a potential very-high-energy-emitting blazar. We present the analysis on the two   other candidates in Appendix~\ref{appendix}.
 
 In Fig.~\ref{fig:0658} we show the \nvssB{} images produced at 4.9 GHz, 8.4 GHz and 23.5 GHz. The parameters of the images are reported in Table~\ref{tab:obs_res}. The source  slightly extends   towards the west direction. 
With the VLBI data we were able to partially resolve the jet structure. To represent the morphology of this emission by means of discrete Gaussian components, we performed the modelfit analysis with the DIFMAP routine. 
In the 4.9 GHz image, the best-fit representation of \nvssB{} is obtained with a point-like core component and two additional components for the jet.  We carried out the modelfit analysis also in the RFC data, finding no evidence of standing-shocks as bright knots present at the different epochs.  The  modelfit best-fit parameters are reported in Table~\ref{tab:mf_0658} in Appendix~\ref{modelfit}. 
 
 \nvssB{} is compact at arcsecond-scales as
deduced from the consistency between the peak brightness (i.e., the core emission) and
the total flux density in the NVSS and VLASS observations (Table~\ref{tab:nvss-first}). Moreover,  the TeV Effelsberg Long-term AGN MONitoring (TELAMON) program  targeted  \nvssB{} after the \icnD{} event. By comparing our results to the TELAMON ones  \citep{Kadler:20212/},   the VLBI total flux density of \nvssB{} at 4.9 GHz seems to be consistent with the   flux density recorded around the same frequency  by the Effelsberg single-dish, suggesting   that the extended emission from this source is negligible and confirming its compact nature.

Table~\ref{tab:ind-spec} reports the spectral indices of the core measured  with the VLBI data.  The spectrum is self-absorbed  between lower frequencies and flat between high frequencies. The same behaviour was observed from the archival RFC data  (Table~\ref{tab:rfc_ind-spec}). 

\begin{figure}
    \centering
    \includegraphics[height=0.45\textwidth]{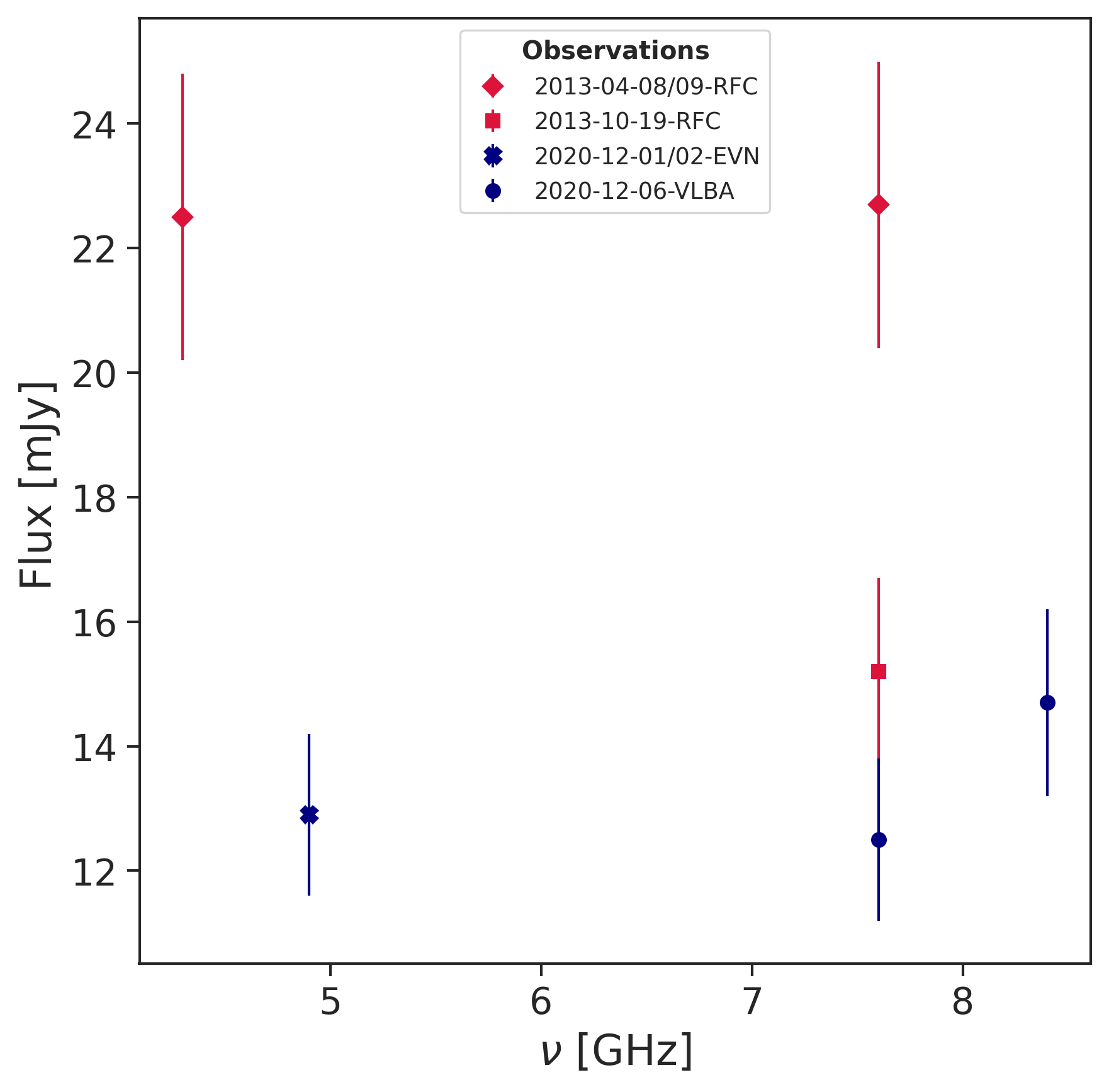}
\caption{   \nvssB{} total radio flux density at different frequencies and epochs.
    \label{fig:0658sp_ind}
}
\end{figure}

Both NVSS and  2017-VLASS data of   \nvssB{}  show a higher flux density  compared to our VLBI results. 
Slightly lower flux density values with respect to the first VLASS epoch are also derived in  the second   VLASS epoch, taken four months before the neutrino event and three years after the first epoch.
Therefore, also archival data  seem to suggest  that in the time range soon before and soon after the neutrino detection,  \nvssB{}  was experiencing   a low activity state  in the radio band (Fig.~\ref{fig:0658sp_ind}).
The decrease of the radio flux density on the mas-scales is up to $V \simeq 0.17$ at around 4.3 GHz  between 2013 and 2020 (Table~\ref{tab:variab}). On the other hand, the TELAMON monitoring of  \nvssB{} \citep{Kadler:20212/} suggests the presence of variability on potentially much smaller time scales.    


\section{Discussion}\label{discussion}

 We conducted VLBI follow-ups of cosmic neutrino events in order to analyse the status of the radio emission of the blazar sources  spatially consistent with  these events.  We presented a total of five sources potentially associated with four IceCube events. We identified one favoured candidate (on the basis of  its blazar-like nature and its association with a $\gamma$-ray source) for each event. Only in the case of \icnC{}  we investigated two sources, one of them  is the only out-of-90\%-neutrino-error-region object studied in detail  in this work. 
 It was included in the analysis   on the basis of its potentially interesting variability and multi-wavelength  characteristics (see Sect. \ref{result}). 
In our search for possible blazar neutrino emitters, we presented a heterogeneous sample of objects. Here we discuss our results on the properties of these candidate counterparts in comparison with the current     knowledge about the neutrino-blazar connection. 

The typical radio properties of the  blazars, such as the core-dominated   jet structures and the flat/inverted spectra, drive our candidates selection among  the radio sources spatially coincident   with the neutrino events that we have observed. These properties 
are confirmed with our VLBI data. Also in the  $\gamma$-ray regime, we found  expected values of $\gamma$-ray photon index (Table~\ref{tab:gammaray}): harder $\gamma$-ray photon indices are observed in   HSP-like sources while softer ones are found for FSRQ sources.

\subsection{On the jet  morphology  and kinematics}

 Theoretical  arguments indicate that   compact parsec-scale regions  of blazars are     site of efficient neutrino production.
 Focused analysis on \txsA{} suggest a connection between this blazar and the neutrino emission. In particular,   the increasing  of the core size of \txsA{} and of the opening angle of its jet  was clearly identified after the neutrino detection. These were   addressed as     observational key features  indicating the neutrino production  \citep{2020A&A...633L...1R}. 
Moreover,  from the evolution of the P.A., \citet{2020ApJ...896...63L} inferred an helical structure in the \txsA{} jet. They pointed out the link between this configuration and the occurrence of instabilities at the  base of the jet. These mechanisms,  in turn, likely drive efficient particles acceleration and   the neutrino production in these sites.
 
All the five  candidates  presented in this work are found to be compact and core-dominated from  mas to arcsecond-scales.
 We could resolve a faint one-sided   jet for three of the candidates: the two FSRQ sources, \txsB{} (associated with \icnB{}) and \pks{} (associated with \icnC{}); and the blazar candidate \nvssB{} (associated with   \icnD{}). 
 In the e-MERLIN image of the faintest source of our sample, \rxs{} (associated with \icnC{}), we observed only a small blob in the south-east direction from the core, while in the HBL object, \whsp{} (associated with   \icnA{}), no  jet structure was  detected in our observations. 
 Performing the modelfit analysis on the jetted sources for which we have multi-frequency observations,  \txsB{} and \nvssB{},   we did not reveal distinct bright knots present in their jets at different frequencies. 
 Also, in our data we did not  identify components  which can be clearly recognized   in the archival RFC    data at the same frequencies.  From the analysis of two epochs of 2016-MOJAVE data, \citet{2019ApJ...874...43L} highlight the presence of a discrete component in the jet of  \pks{}.   They measured an apparent speed of this component of $(12\pm8)$ $\mu$as\,yr$^{-1}$, that is (0.40$\pm$0.28)c.  
 Similar value  was found by \citet{2020ApJ...896...63L} for the closest-to-the-core   component  of the \txsA{} jet. Testing the kinematic and the evolution  of this component with future high resolution VLBI data could  become relevant  in order to compare the \pks{} and \txsA{} cases and infer the possible relation  with the neutrino production.  
 
  The   trend  of the  P.A. of the \txsB{}, \pks{}   and \nvssB{} jets  seems to remain unchanged over the period between the observations (see Table~\ref{tab:mf_txs}    for \txsB{}, Table~\ref{tab:mf_pks}  for  \pks{} and Table~\ref{tab:mf_0658}   for \nvssB{}). 
  The analysis of the P.A. variation with time performed by \citet{2020ApJ...896...63L} is not   fully compatible  to the cases here studied      because we could  not identify   the comparable jet components in each epoch and because the data are poorly sampled in time.

 As far as the relation between neutrino production and blazar core size is concerned, in the case of \txsA{}, the source increased its core size from 68 $\mu$as to 158 $\mu$as  in the six months after the neutrino detection  \citep{2020A&A...633L...1R}. The apparent expansion occurred at about twice the speed of light at the redshift of \txsA{} \citep[$z=0.34$,][]{2018ApJ...854L..32P}. As suggested by \citet{2020A&A...633L...1R},  this behavior should test the presence of ongoing hadronic processes of which high-energy neutrinos are very likely products. In the case of \txsB{}, the 8.4 GHz  core  is  smaller in our  observations  than in the 2007 and 2012 RFC observations at the same frequency (see  Table~\ref{tab:mf_txs}). However,   the core linear size  is much larger  in \txsB{}  (about 1.2 pc)  than in  \txsA{}, both before ($\sim$0.33 pc) and after   ($\sim$0.76 pc) the apparent superluminal expansion; this is due to  \txsB{} being located at redshift 0.91 while \txsA{} at redshfit 0.34. Based on  the RFC and our data, we can   speculate  that \txsB{} is in its initial compact phase, which could be followed by an expansion phase in the future. 

 In the case of \pks{}, throughout 2020 MOJAVE  observations (15.3 GHz), the core size  did not change significantly within the uncertainties (see  Table~\ref{tab:mf_pks}). Also in this case, however, the source is more distant that \txsA{} and therefore our angular resolution does not allow us to probe the scales of the expansion discussed by  \citet{2020A&A...633L...1R}.
 An important caveat in this context is  that  \citet{2020A&A...633L...1R} deduced the superluminal expansion of the core from 43 GHz VLBA observations. Due to the small beam size,  mm-VLBA  observations are  a desirable approach for future core size expansion measurements.


 We derived the observed    core brightness temperatures, $T_{B}^{obs}$,  by using Equation (2) from \citet{1982ApJ...252..102C}. For \txsB{}, we found $T_{B}^{obs}= 4.1\times 10^{11}$ K in the 2007 RFC data, $T_{B}^{obs}=8.6\times 10^{11}$ in the 2012 RFC data and  $T_{B}^{obs}=7.7\times 10^{12}$ K  in our 2020 data, all at 8.4 GHz. The brightness temperature depends not only on the core geometric factors but also on its flux density.  Differences in brightness temperature values reflect then different physical conditions in the core.   The observed variation might be a result of a larger Doppler boosting factor in the 2020 data compared to the past \citep[e.g.,][]{2000ARep...44..719K,2002PASA...19...77K}. In general, such values of $T_{B}^{obs}$ suggest the presence of a highly relativistic flow.

 The core brightness temperature of \pks{} at 15.3 GHz is   $T_{B}^{obs}=20.2\times 10^{12}$ K based on the May 2020 data,  
 $T_{B}^{obs}=6.0\times 10^{12}$ K in the October 2020 data, and $T_{B}^{obs}=20.3\times 10^{12}$ K in the December 2020 data.

 At 15.3 GHz, \citet{2020ApJ...896...63L} observed values of $T_{B}$ in \txsA{} core which are  significantly higher in the observing epochs following the \icntxs{} event with respect to the previous observations. However, it must be noted that
   a rigorous comparison between the  modelfit results from our and RFC  data is not actually possible   because of the difference between the ($uv$)-sampling of our and RFC data. For this reason  also the core component of both \txsB{} and  \pks{}  is not unambiguously and uniformly identified in each epoch.

Limb-brightening  and transverse structure   features are invoked for explaining the high-energy neutrino emission when an external (to the jet) seed photon field is lacking.  This is the case of BLL objects for which a two-layer spine-sheath jet  have been proposed     \citep{2014ApJ...793L..18T}. The two layers of the jet, moving with  different Doppler factors (highly relativistic the spine and slower the external sheath),  are thought to provide an energetic enough photon field  owing to  their relative  motion. 
In this context,  limb-brightening  and transverse structure  are then expected to be observed in possible neutrino counterparts and, indeed, these have been found by \citet{2020A&A...633L...1R}  in \txsA{}.  Interestingly,   \pksa{} jet  \citep[proposed as counterpart of HESE-35;][]{Kadler2016} also exhibits hints of transverse  structures  in its tapered image  \citep{2010A&A...519A..45O}.  
 Based on a simple, by-eye inspection of the flux density contour levels of the images, limb-brightening configuration are not visible in any of the jetted sources. 
In the case of  \nvssB{},  if the    spine-sheath layers structure is present,     the   low flux density of the source combined to the observing limitations could have prevented the  detection of such features.  

\subsection{On the flux density variability}

A temporal connection between an increased flux density around 22 GHz and the neutrino arrival is deduced in VLBI observation by \citet{2020ApJ...894..101P}, and confirmed with single-dish observations 15 GHz by \citet{2021A&A...650A..83H} for a sample of
blazars. Single-source works also support this idea
\citep[e.g.,][]{Kadler2016,2020A&A...633L...1R}. 
According to \citet{2020ApJ...894..101P},  the simultaneity between the neutrino event and the radio flare could be due to an energetic radio outflow arising from particle injection mechanisms occurring near the central black hole. Based on the energetic requirements and the synchrotron opacity constraints, \citet{2020ApJ...894..101P,2021ApJ...908..157P}  were also able to confine the region of neutrinos origin in the core of blazars, within a few parsecs at the jet-base.  
The increased radio activity in \txsA{} can be  noted in the light curves reported by \citet{IceCube2018a}, and it results in a significant \citep[according to ][]{1992ApJ...399...16A}  variability of $V\sim0.1$  from the VLBI data referred to the time range  between 2017-11 and 2018-05 \citep[these values of flux density   are taken from ][]{2020A&A...633L...1R}.  
We highlight that the variability index defined in Eq.~\ref{eq:var} is mostly useful to compare the flux density variations in objects which show large differences in flux density levels. In this way, the $V$-index allows us to quantitatively verify if the emission of a source  has significantly varied.   
However,  it could be not sufficient to catch the magnitude of radio flares, as can be seen in the low value of $V$ measured even for the case of \txsA{}.
The same consideration holds with other variability indicators.   \citet{2021A&A...650A..83H} applied an activity index to the OVRO light curves data of a sample of blazars to detect the presence of high states.
They claim that also their approach could fail to properly reflect notable flaring states of the sources.  \citet{2020ApJ...894..101P} point out that the self-absorption in the more compact (and then variable) central regions,  together with the contribution of the unresolved diffuse emission,  could lead to underestimating the variability in the radio activity of the sources.

In our sample, one of the five sources,  \pks{}, shows a  high state of activity in the radio band.  
We revealed an increased activity ($V\sim0.1$) from the 15.3 GHz MOJAVE data of this source. The neutrino was detected on October 21, 2020, while the 15.3 GHz flux density started to grow from May 2020 (Table~\ref{tab:rfc}).  

The preliminary results from the RATAN-600 observations, reported by \citet{2020ATel13405....1K}, suggest an increase of the \txsB{} emission soon after the neutrino detection and before our observations (carried out one month after the neutrino arrival).  At the same RATAN-600 frequency we observed a lower value of flux density.   As discussed in Sect.~\ref{result}, the discrepancy between our and RATAN-600 flux density holds  also if we account for the VLBA calibration issues described in the NRAO notice.    The discrepancy can be due to either the different sensitivity of RATAN-600 and VLBA to the diffuse emission, or to the variability of the source itself.

 An extraordinary low activity level of \txsB{} was   observed in the 2012 RFC data (Table~\ref{tab:rfc}). If we compare our flux density measurement to this observation we could speculate, in line with RATAN-600 results, that the source is in fact in a significantly higher state also in our observation (Table~\ref{tab:variab}). However,  being this RFC flux density measurement from a single isolated    epoch, separated by several years from our observation, the indication of an increased flux density does not lead to firm conclusions. In our VLBI data,  \txsB{} was in a  state of activity compatible with the 2004 and 2007 RFC  observations.  

 While \whsp{}   does not seem to exhibit an enhanced flux density and no VLBI archival data of   \rxs{} are available to check its variability, the last of the five sources, \nvssB{},  is clearly in a low state of activity in the radio band at the neutrino arrival, as deduced by the comparison of our results with the archival RFC observations (Table~\ref{tab:variab}). The largest change in the flux density follows by comparing the 2013 RFC data with our EVN observation, both carried out at $\sim$4.3 GHz. As noted above and reported in several works,  the low frequencies are less affected by the activity occurring within the jet base - core (i.e. the region in which is thought to happen the neutrino production).  
Since we miss archival data of \nvssB{} at high frequency,  we could not determine the state of activity at these frequencies.
However, the lack of an enhanced state of activity is not in contrast to what was found in the case of \txsA{}. Indeed, after the \icntxs{} event, the inspection of archival IceCube and multi-wavelength data from the direction of \txsA{}   resulted in evidence of a neutrino flux excess from that position. In coincidence with this archival neutrino detection, no radio flares have been detected, as well as other-wavelength flares.

\subsection{High-energy neutrino production}

The origin of neutrinos in blazars can be associated with regions enabling the acceleration of relativistic protons and cosmic rays \cite[e.g.,][]{2020NewAR..8901543M}. These sites can be hosted by the accretion disk and relativistic jet which offer a favourable environment for hadronic and photo-hadronic interactions that produce neutrinos \cite[e.g.,][]{1995APh.....3..295M,2020NewAR..8901543M,2019MNRAS.483L.127R,2020ApJ...902..108M}. The shock acceleration of protons in a turbulent environment must occur over timescales shorter than that involving synchrotron energy loss or the diffusion of protons away from the acceleration zone \cite[e.g.,][]{2017PhRvD..96f3007Z}. 

The shock itself may originate from a diversity of physical processes in the central nuclear region. 
Jet scenarios include through propagating or re-confinement shocks at varying distances from the supermassive black hole \cite[e.g.,][]{2002A&A...386..833G,2020NewAR..8901543M,2021A&A...654A..96Z}, magnetic re-connections \cite[e.g.,][]{2020NatCo..11.4176S,2020NewAR..8901543M}, interaction with transiting gravitationally bound clouds or clumps \cite[e.g.,][]{2010A&A...522A..97A}, or from an interaction with the external radiation field \cite[e.g.,][]{2020MNRAS.496.2885H,2021NatAs...5..472W}.

The physical characteristics of the production region can be probed by comparing the acceleration timescale for protons $t_{\rm acc}$ with the dynamical timescale over which any changes are propagated, $t_{\rm dyn}$. Assuming a second-order Fermi acceleration of the protons \cite[e.g.,][]{2015ApJ...806..159K,2019ApJ...886..114H},
\begin{equation}
t_{\rm acc} \approx 10 \frac{r}{c} \left(\frac{v_A}{c}\right)^{-2} \left(\frac{r_L}{r}\right)^{2-s} \gamma^{2-s}, 
\end{equation} 
where $r$ is the radial distance to the acceleration location, $v_A = B/(4 \pi \rho)^{1/2}$ is the Alfv\'en velocity expressed in terms of the magnetic field strength $B$ and local density $\rho$, $r_L \sim m_p c^2/(e B)$ and $\gamma$ are the Larmor radius and the Lorentz factor of the protons, and    $s = 1.5 - 2$ is the spectral index of the turbulence scale length distribution.

For the jet scenario, the dynamical timescale $t_{\rm dyn}$ is the typical time for the propagation of a signal in a region of size $\varpi = r/\Gamma$ projected along the observer line of sight (assuming a jet bulk Lorentz factor $\Gamma$) and is
\begin{equation}
t_{\rm dyn} = \frac{r}{v_j} = \frac{\tilde{r} R_S}{v_j},
\end{equation}
where $v_j/c = (1-\Gamma^{-2})^{1/2}$ is the jet bulk velocity scaled in units of $c$,  $R_S = 2 G M_\bullet/c^2$ is the Schwarzschild radius around a black hole of mass $M_\bullet$, and   $\tilde{r} = r/R_S$ is a scaled radial distance. The density of plasma in the jet can be evaluated using

\begin{equation}
\rho = \frac{L_{\rm j,kin}}{\frac{\pi r^2}{2} v^3_j c^3 \Gamma^2},
\end{equation}
where $L_{\rm j,kin}$ is the kinetic energy in the jet. Assuming that the energy equipartition holds in the jet, the magnetic field strength is

\begin{equation}
B = \left(\frac{8 \pi}{3 \beta}\right)^{1/2} (\rho v^2_j)^{1/2} = \frac{4}{r \Gamma} \left(\frac{L_{\rm j,kin}}{3 \beta v_j}\right)^{1/2}.
\end{equation}
where $\beta$ is the plasma beta (ratio between the plasma pressure
and the magnetic field pressure). 
With the condition $t_{\rm acc} \leqslant t_{\rm dyn}$, assuming $\beta = 1$, $s = 1.5$, $v_j/c \approx 1$,
\begin{equation}
E_p = \gamma m_p c^2  \leqslant (7.11 \times 10^{16}~{\rm eV})~\left(\frac{L_{\rm j,kin}}{10^{46}~{\rm erg}~{\rm s}^{-1}}\right)^{1/2} \left(\frac{\Gamma}{5}\right)^{-2}.
\end{equation}

The above rough estimates indicate that protons can certainly be accelerated to PeV or greater energies. Individual neutrinos up to PeV energies can be produced with the availability of suitable cross sections for the hadronic and photo-hadronic interactions as they are expected to have energies $E_\nu \approx E_p/20$ \cite[e.g.,][]{2019ApJ...886..114H}. This toy model then offers strong
support of jet as host of energetic protons and neutrinos. Distinguishing between this and other scenarios, as the one involving mechanism occurring at the accretion disk site, is beyond the scope of the current work. 

\subsection{Searching  for neutrino emitting sources}

Other neutrino emitter candidates are currently taken into account. One of the most intriguing phenomena is the case of the tidal disruption event (TDE)  AT2019dsg,  associated with high probability with the IC\,191001A detection  \citep{2021NatAs...5..510S}. TDEs take place when a star orbits too close to the central supermassive black hole in a galaxy and it is destroyed under the action of the tidal force \citep[e.g.,][for a review]{2013IAUS..290...53K}. The star destruption releases energetic outflows able to produce high-energy cosmic particles and high-energy neutrinos.
 Other studies are mostly focusing on the connection between the X-ray emission from blazar and the neutrino events  \citep[e.g.,][]{Stathopoulos:2021TR}, while others are investigating specific classes of AGN \citep[e.g.,][]{2017ICRC...35.1000M} or all radio-loud AGN \citep{Larson:2021NX}. According to
 \citet{2020MNRAS.497..865G},    IBL/HBLs are promising counterparts of neutrinos.

The growing number of astrophysical objects showing hints of a connection with neutrino emission suggests we should keep the door open to all kinds of possible neutrino event-related sources. In this context, we also observed non-blazar-like sources (reported in Appendix~\ref{appendix}) and sources slightly out of the neutrino error region to leave an observational reference for further studies on the astrophysical neutrino counterparts. The sources analysed in this work have not been previously studied in detail and for some of them, we provide also   VLBI observations for the first time. We stress the fact that our VLBI observations are crucial,  not only for the improved quality of the data with respect to the archival VLBI data of the targets but, above all, for their  \textit{temporal coincidence} with the (still not easily detectable) neutrino arrivals. Then, these data give a unique opportunity to investigate the evolution of the possible electromagnetic counterparts after the neutrino detection.

In particular, dense monitoring of the sources is required to efficiently estimate proper motions of the components of the jet \citep[e.g.,][]{2013A&A...559A..75B}.  Proper motion studies will allow us to infer the physical and geometrical parameters of the jets.  As shown in the previous Section, some of these parameters, as, for example, the magnetic field strength or the jet bulk Lorentz factor, could provide a test for
the theoretical expectations.

Due to the limited sensitivity of IceCube at high energies, the collected detections are reasonably the tip of the iceberg of larger low-energy neutrino flux. In this regard, retrieving the IceCube archival data of the neutrinos observed at lower energies could be worthwhile to catch the presence of low-energy neutrino excess from the position of the sources here analysed.
The IceCube Realtime Streams do not provide all the high-energy events.

So far, no source has been found coincident with a significant excess of neutrinos, except for the known cases of evidence found for \txsA{} and NGC\,1068 \citep{2020PhRvL.124e1103A}.
These kinds of searches are often penalized by a large number of trial factors and it would be important to have a limited number of source positions be tested based on their potential association with neutrino events. Therefore, we suggest performing dedicated analysis of neutrino data testing detections at the position of our candidates  \pks{}, \txsB{} and \nvssB{} for which the connection with the neutrino production is supported by some observational indications in our paper. This could lead, in an optimistic scenario, to findings similar to 2014--2015 neutrino excess from the direction of \txsA{}.

\section{Summary and conclusions}\label{concl}

We presented VLBI follow-up observing results to four IceCube neutrino events, both with high and moderate (Gold and Bronze) probability to be associated with a  cosmic origin. We analysed data of a total of ten radio sources in spatial coincidence with the neutrino events. Following previous results reported for the case of \txsA{}, and the ones based on statistical approaches using VLBI data, we aimed to identify the possible neutrino emitters by testing the presence of radio properties connected to the neutrino production processes. We used our follow-up data in comparison to VLA (NVSS, FIRST, VLASS) surveys and archival VLBI (RFC) data. The candidate neutrino counterparts have a heterogeneous optical classification, ranging from BLL to FSRQ and including a few sources of uncertain classification. Among the ten candidates, we selected five blazar-like sources (on the base of their radio properties) with a $\gamma$-ray counterpart. A description of the other candidate counterparts is provided in Appendix~\ref{appendix}.  The main outcomes on the five principal candidates can be summarized as follows:

\begin{itemize}

    \item  The core spectral behavior of all the sources is well represented by a  self-absorbed spectrum at low radio frequencies and a flat spectrum at high radio frequencies. When the objects are bright and the image noise allowed it, we were able to recognize a core-jet morphology.  These appear in three of the sources. All sources are compact from mas VLBI resolutions to arcsecond VLA surveys resolutions.
    
    \item From a morphological point of view,
    we did not identify parsec-scale radio properties in our sample that could be linked to the neutrino emission, as it was previously studied in the case of \txsA{} and expected from theoretical arguments.   This can be mostly due to the lower quality of archival data compared to our new data and the lack of frequent observations of these sources in the past at  VLBI  resolution.

 \item  One of the five candidates, the FSRQ \pks{},  exhibits hints of an enhanced activity state in our data. However, it is found outside the 90\% localisation region around the neutrino position. Another source, \nvssB{}, is found in a low radio activity state.  The case of \txsB{} remains ambiguous since in only one archival RFC observation this source shows a lower flux density with respect to our results,  while previous RFC flux density measurements are in agreement with our results.  On the other hand,  \citet{2020ATel13405....1K}  confirm  a high state of  \txsB{} at the neutrino arrival.

\item None of the sources show exceptionally outstanding properties, in terms of their radio luminosity, variability, and kinematics. However, while we are well able to derive the radio luminosity values from our observations, the variability and the kinematics parameters are much less constrained. 

 \item We present a toy model for proton acceleration in jets which supports the jet scenario for  neutrino production in these sites. The model requires kinetic parameters that can be inferred by well  time-sampled VLBI monitoring of jets. As mentioned above, a study of the kinetic parameters of the jet motion will require longer, dedicated VLBI monitoring campaign that we are undertaking as a development of this project.

\end{itemize}

In conclusion, based on the  analysis of the morphological evolution, we cannot exclude or support the hypothesis of the sources analysed as a potential neutrino counterpart. However, the results which suggest significant variability in the radio band in  \pks{} at the neutrino arrival are consistent with the evidence of radio flares in blazars in temporal coincidence with neutrino emission \citep{2020ApJ...894..101P,2021A&A...650A..83H}.

 VLBI results provide us with important pieces of information on the neutrino candidates. In addition, a complete characterization of these candidate neutrino-associated sources could hopefully come from the combined efforts of multi-wavelength campaigns, triggered almost simultaneously to our follow-up \citep[e.g.,][de Menezes et al. in prep]{deMenezes:2021Fl}, together with the planned VLBI observations that we will present in future works.  
Further investigation on these candidates and, most importantly, the identification of a  large number of new ones    will shed light on the open question of the nature of extragalactic neutrinos and of blazars as the best candidate neutrino counterparts.

\begin{acknowledgements}
 We thank the referee for her/his useful suggestions. JM and MPT acknowledge financial support from the State Agency for Research of the Spanish MCIU through the
"Center of Excellence Severo Ochoa" award to the Instituto de Astrofísica de Andalucía (SEV-2017-0709)
and through grants RTI2018-096228-B-C31 and PID2020-117404GB-C21 (MICIU/FEDER, EU).   SB acknowledges financial support by the European Research Council for the ERC Starting grant \emph{MessMapp}, under contract no. 949555. BWS is grateful for the support by the National Research Foundation of Korea (NRF) funded by the Ministry of Science and ICT (MSIT) of Korea (NRF-2020K1A3A1A78114060). We thank  to L. Petrov for granting permission for using    data from the Astrogeo VLBI FITS image database.
The European VLBI Network is a joint facility of independent European, African, Asian, and North American radio astronomy institutes. Scientific results from data presented in this publication are derived from the following EVN project codes: RG011, EG108. The National Radio Astronomy Observatory is a facility of the National Science Foundation operated under cooperative agreement by Associated Universities, Inc. This work made use of the Swinburne University of Technology software correlator, developed as part of the Australian Major National Research Facilities Programme and operated under licence. 
e-MERLIN is a National Facility operated by the University of Manchester at Jodrell Bank Observatory on behalf of STFC.    This research has made use of data from the MOJAVE database that is maintained by the MOJAVE team (Lister et al. 2018).
This research has made use of the CIRADA cutout service at URL cutouts.cirada.ca, operated by the Canadian Initiative for Radio Astronomy Data Analysis (CIRADA). CIRADA is funded by a grant from the Canada Foundation for Innovation 2017 Innovation Fund (Project 35999), as well as by the Provinces of Ontario, British Columbia, Alberta, Manitoba and Quebec, in collaboration with the National Research Council of Canada, the US National Radio Astronomy Observatory and Australia’s Commonwealth Scientific and Industrial Research Organisation. 
This research has made use of the NASA/IPAC Extragalactic Database (NED),
which is operated by the Jet Propulsion Laboratory, California Institute of Technology,
under contract with the National Aeronautics and Space Administration. 

\end{acknowledgements}

\bibliographystyle{aa}
\bibliography{2022_03}

\appendix

\section{Other candidates}\label{appendix}

 \paragraph{\textbf{\nvssA{}}}
Lying within the 50\% uncertainty region,  4FGL\,\gammajtfn{} was the first identified candidate source associated with the \icnA{} neutrino event.  Its possible radio counterpart is NVSS J104938+274212 (SDSS\,J104938.80+274213.0, a galaxy at $z$ = 0.144, \citet{2015ApJS..219...12A}).    Parameters of VLBI observation are reported in Table~\ref{tab:obs_res_appendix} and spectral indices deduced from these data are in Table~\ref{tab:ind-spec_app}. The  FIRST 1.4 GHz peak flux density is 9.2 mJy, and the integrated flux density is 15.6 mJy, suggesting that \nvssA{} contains a resolved radio structure on the arcsecond-scales (see Table~\ref{tab:nvss-first}). Our VLBI observations reveal a compact component on parsec-scales, with a lower flux density than  what is observed in the VLA surveys.  The luminosity of the source at 1.5 GHz is (2.1$\pm$0.2)$\times10^{23}$ $\mathrm{W Hz^{-1}}$, assuming a spectral index $\alpha_{\mathrm{NVSS}}^{\mathrm{VLASS}}=-0.3\pm0.2$.

 \paragraph{\textbf{\icnBone{}}}
One of the candidate  $\gamma$-ray counterparts associated with \icnB{} is 4FGL\,J1114.6+1225 (Table~\ref{tab:gammaray}). A possible counterpart of this source is \icnBone{}, the properties of which are reported in Table~\ref{tab:counterpart}.  Among all the candidate counterparts associated with the events of our sample, this is the source located farthest to the position of the relative neutrino (at about  4 degrees away from   \icnB{}). This first leads us to disfavor it as the most promising candidate. Moreover, it is not cataloged as a blazar.  We identified two radio counterparts in spatial coincidence with \icnBone{}. In the following we refer to them with \icnBone{}-A and \icnBone{}-B. Information about their  arcsecond-scale low-frequency properties are provided in Table~\ref{tab:nvss-first}. 

\paragraph{\textit{\icnBone{}-A}} Inspecting the highest resolution image of this source produced at 23.5 GHz, we resolved  two components separated by about 8.4 mas. The second component was also detected in the 8.4 GHz images while it is not detected in the 4.9 GHz image. In Table~\ref{tab:obs_res_appendix}  the properties of the VLBI observations of these components are reported.  
We use the suffix A1 to the brightest and A2 to the less bright detection.  The spectral index between 8.4 GHz and 23.5 GHz of the A1 component is  1.4$\pm$0.1 while the one of the A2 component is $-$1.4$\pm$0.4. From these results, we could speculate that we are looking at a self-absorbed core component (A1) and a contribution from a  steep spectrum jet (A2). However, it must be noted that with these data we were not able to measure the spectral index adequately, that is by adopting the same $uv$-range, pixel size, and restoring beam in the two frequency images. Then the resulting spectral indices could be biased by the differences in the observation setups.

\paragraph{\textit{\icnBone{}-B}}
This source was not detected in our  8.4 GHz and 23.5 GHz observations. The upper limits set as 3 times the rms noise levels of the images are 2.3 mJy and   80 $\mu$Jy at 8.4 GHz and 23.5 GHz respectively.  At 4.9 GHz the source is composed of two components (B1 and B2), located at a distance of around 7 mas from each other.  The properties of these components are reported in  Table~\ref{tab:obs_res_appendix}. Comparing these results to the low-frequency ones (Table~\ref{tab:nvss-first}),  the source seems to be characterised by a steep spectrum.

 \paragraph{\textbf{\icnBtwo{}}} 
 The other  source within the error region  of the \icnB{} event is \icnBtwo{} (Table~\ref{tab:counterpart}). The possible radio counterpart observed by the VLBI shows a significant emission outside the core, as deduced by comparing the integrated flux density and the peak intensity at 4.9 GHz and 8.4 GHz (Table~\ref{tab:obs_res_appendix}).  In our 23.5 GHz data, \icnBtwo{} lies under the rms noise level of the image, i.e., 0.9 $\mathrm{mJy\,beam^{-1}}$. The corresponding upper limit for the surface brightness of the source is 2.7 $\mathrm{mJy\,beam^{-1}}$.  The NVSS, FIRST, and VLASS data reveal that the large fraction of the source emission is spread over arcsecond-scales   (Table~\ref{tab:nvss-first}). This is also confirmed by the steep spectral index  obtained from our 4.9 GHz and 8.4 GHz data (Table~\ref{tab:ind-spec_app}). 

 \paragraph{\textbf{\wisea{}}}
The blazar-like source \wisea{} is one of the possible   \icnD{} counterparts  \citep{2020ATel14225....1G}. However, both the absence of a $\gamma$-ray association with this source and its position outside the 90\% error region of the neutrino event (see Table~\ref{tab:neutrino}), lead us to disfavour  \wisea{} as the best candidate. 
In the VLBI observations (Table~\ref{tab:obs_res_appendix}), the source shows a partially resolved structure with a core and a shallow emission extended towards the southwest direction. Due to the large beam size of the NVSS data, it results that the source emission is embedded with a close ($\sim$44$''$) source in the field. In the VLASS these two objects are separated. We measured the two contributions in NVSS data by fitting the object's emission with two Gaussian components using the \texttt{imfit} tool of CASA.  For completeness, we report the parameters of both the sources, labeled \wisea{}-A and \wisea{}-B in Table~\ref{tab:nvss-first}. The target of our observation is  \wisea{}-A. The contribution of \wisea{}-B is not detected in our VLBI data.   Table~\ref{tab:ind-spec_app} reports the VLBI spectral indices of the main target source. We confirm the blazar nature of this source, based on the large core prominence (as indicated by the comparison of the VLBI and NVSS/VLASS flux densities) and the behaviour of the spectral index, which is inverted at low frequency and flat at higher frequencies.

 \paragraph{\textbf{\nvssC{}}}
The second blazar candidate associated with  \icnD{} is \nvssC{}  \citep{2020ATel14225....1G}.  This source is not associated with any $\gamma$-ray detection, either  (Table~\ref{tab:counterpart}). In our VLBI observations, \nvssC{} is composed by a bright core and an elongated jet that extends in the north-west direction.
In  Table~\ref{tab:obs_res_appendix} we report  VLBI properties of this source. Surveys data show that this source is core-dominated at arcsecond-scales  (Table~\ref{tab:nvss-first}). The VLASS flux densities in the two epochs (2017 and 2020, two months before the neutrino detection) are consistent within the errors, implying the absence of variability at VLASS-scales in that time range. In Table~\ref{tab:ind-spec_app} we report  the spectral index measurements.

\begin{table*}[htb]
\centering
\small
\begin{threeparttable}
\caption{Images parameters of VLBI observations.}
 \begin{tabular}{l c c c c c }
 \hline \hline
Source &  $\nu$ & $S_{\mathrm{peak}}$ & {$S_{\mathrm{int}}$}       & rms                           &Beam\\ 
{}     &  $\mathrm{(GHz)}$ &$\mathrm{(mJy\,beam^{-1})}$&{$\mathrm{(mJy)}$}&{$\mathrm{(\mu Jy\,beam^{-1})}$} &({mas$\times$mas,\deg})\\
\small(1) & \small(2) &\small (3) &\small (4) &\small (5) &\small (6) \\
\hline 
\textbf{190704A}\\
\small{\nvssA{}} &  1.5   & 3.0$\pm$0.3 & 4.9$\pm$0.5 & 83 & 11.6$\times$6.2, $-$24.5\\
                 &  4.4   & 5.7$\pm$0.6 & 6.3$\pm$0.6 & 136  & 5.7$\times$5.3, 23.1\\
                 &  7.6   & 3.3$\pm$0.5 & 4.8$\pm$0.6 & 143  & 2.4$\times$1.3, $-$20.1\\
\hline
\textbf{200109A} \\
\small{\icnBone{}-A} & 4.9   &  1.0$\pm$0.1 &  1.7$\pm$0.3 & 57 & 1.8$\times$1.5,  45.0  \\
\,\,\,\,\,\,\,\,\,\,\,\, \,\,\,\,\,\,\,\,\,\,\,\, \,\,\,\,\,\,\,\,\,\,\,\, \,\,\,\,\,\,\,\,\,\,\,\, \,\,\,\,\,\,\,\,\,\,\,\, \,\,\,\,  \small{A1}  &   8.4   &  1.8$\pm$0.2 &   2.2$\pm$0.2 & 66  &  2.2$\times$1.0, $-$5.5 \\
                 &  23.5 & 0.30$\pm$0.03  & 0.55$\pm$0.07  & 95  &  1.2$\times$0.8, $-$8.1 \\
\,\,\,\,\,\,\,\,\,\,\,\, \,\,\,\,\,\,\,\,\,\,\,\, \,\,\,\,\,\,\,\,\,\,\,\, \,\,\,\,\,\,\,\,\,\,\,\, \,\,\,\,\,\,\,\,\,\,\,\, \,\,\,\, \small{A2} &   8.4   &  0.25$\pm$0.05 & 0.14$\pm$0.05    &    66  &  2.2$\times$1.0, $-$5.5 \\
                 &  23.5 & 0.42$\pm$0.05  & 0.64$\pm$0.08  & 95  &  1.2$\times$0.8, $-$8.1  \\
                 \hline
\small{\icnBone{}-B1} & 4.9   &  0.39$\pm$0.05 &  0.52$\pm$0.08 & 34 & 1.8$\times$1.5,  30.6  \\
\,\,\,\,\,\,\,\,\,\,\,\, \,\,\,\,\,\,\,\,\,\,\,\, 
\,\,\,\,\,\,\,\,\,\,\,\, \,\,\,\,\,\,\,\,\,\,\,\, 
\,\,\,\,\,\,\,\,\,\,\,\, 
\,\,\,\, \,\, \small{B2}&  &  0.22$\pm$0.03 &  0.41$\pm$0.07 &    &    \\
                 \hline
\small{\icnBtwo{}} &  4.9  & 17.2$\pm$1.8 &  34.1$\pm$4.1  & 476 & 2.6$\times$2.2, 79.0\\
                   &  8.4  &  8.4$\pm$0.9 &  24.4$\pm$4.2  &  452  &  2.8$\times$2.0, 0.8  \\
                 \hline
\textbf{201114A} \\
\small{\wisea{}} & 4.9 & 25.3$\pm$2.7 & 38.6$\pm$4.3 & 58 & 3.5$\times$1.7, 74.5\\
                 & 8.4 & 68.7$\pm$6.9 & 76.0$\pm$7.6 & 95 & 2.1$\times$1.1, 1.1 \\
                 & 23.5& 47.4$\pm$4.7 & 49.3$\pm$4.9 & 96 & 0.9$\times$0.4, $-$9.9 \\
                 \hline
\small{\nvssC{}} & 4.9 & 521$\pm$52   & 734$\pm$74 & 262  & 2.8$\times$2.0 73.2 \\
                 & 8.4 & 421$\pm$42 & 568$\pm$57 & 254 & 2.3$\times$1.1, $-$7.1 \\
                 & 23.5& 129$\pm$13 & 192$\pm$21 &  766 &  1.1$\times$0.4, $-$14.4\\ 
                          \hline

 \end{tabular}
\begin{tablenotes}[para,flushleft]
\item 
 \textit{Notes:}
Col. 1 -  Candidate neutrino counterpart; 
Col. 2 -  Observation Frequency in GHz; 
Col. 3 -  Peak brightness in $\mathrm{mJy\,beam^{-1}}$;
Col. 4 -  Integrated flux density in mJy;
Col. 5 -  1-$\sigma$  noise level of the image in $\mathrm{\mu Jy\,beam^{-1}}$; 
Col. 6 -  Major axis (in mas), minor axis (in mas), and P.A.  (in degrees, measured from North to East)  of the restoring beam.
Parameters are referred to  natural weighting images.
\end{tablenotes}
	\label{tab:obs_res_appendix}
\end{threeparttable}
\end{table*}

\begin{table*}[htb]
\centering
\small
\begin{threeparttable}
\caption{Spectral index measured with VLBI data.}
 \begin{tabular}{c c c c c c c c c }
 \hline \hline
IC name & Source & $\nu$            & $S_{\mathrm{peak}}$ & $uv$-range & Beam &  $\alpha$  \\ 
         &        &  $\mathrm{(GHz)}$ & $\mathrm{(mJy\,beam^{-1})}$& M$\lambda$ & {mas$\times$mas,\deg}& \\
\small(1) & \small(2) &\small (3) &\small (4) &\small (5) &\small (6) & \small(7)  \\
\hline  
\textbf{190704A} & \nvssA{} & 1.5 & 2.9$\pm$0.3 & 2-40  & 6.6$\times$5.6, 67.0      &  0.7$\pm$0.1   \\
                 &            &  4.4 & 5.9$\pm$0.6 &       &                           & \\
\hline
                 &           &   4.4 & 3.8$\pm$0.4 & 5-100 & 2.8$\times$1.7, $-$5.3 &  0.0$\pm$0.3 \\
                 &           &   7.6 & 3.8$\pm$0.4 &       &                           &  \\
                   \hline
\textbf{200109A} & \icnBtwo{} &  4.9 & 16.3$\pm$1.7 & 4-155  &  2.6$\times$1.9, 0.8      &  $-$1.4$\pm$0.3   \\
                 &            &  8.4 &  7.8$\pm$0.9 &       &                           & \\
\hline
 \textbf{201114A}  & \wisea{} & 4.9 & 25.1$\pm$2.5  & 13-144 & 1.5$\times$1.1, 6.4 &  1.8$\pm$0.3  \\
                &            &  8.4 & 67.6$\pm$6.8  &   &  & \\
\hline
                &            & 8.4  & 66.6$\pm$6.6 & 13-244 & 1.4$\times$0.9, 0.9 &  $-$0.3$\pm$0.1 \\
                &            &    23.5  & 49.1$\pm$4.9 &  & & \\
\hline 
              &    \nvssC{} & 4.9    & 421$\pm$42 & 8-165   & 2.2$\times$1.2, $-$7.5  &  $-$0.4$\pm$0.3  \\
                &            &  8.4 &  520$\pm$52    &   &  & \\
\hline
                &            & 8.4  & 385$\pm$39 & 12-245 & 1.6$\times$0.9, $-$15.0 &  $-$0.9$\pm$0.1 \\
                &            & 23.5 & 154$\pm$15  &        &                          & \\
                         \hline

 \end{tabular}
\begin{tablenotes}[para,flushleft]
\item 
 \textit{Notes:}
Col. 1 -  IceCube event name; 
Col. 2 -  Candidate neutrino counterpart;
Col. 3 -  Observation frequency in GHz; 
Col. 4 -  Peak intensity in $\mathrm{mJy\,beam^{-1}}$;
Col. 5 -   $uv$-range in M$\lambda$;
Col. 6 -  Beam sizes;
Col. 7 -  Spectral index.
\end{tablenotes}
	\label{tab:ind-spec_app}
\end{threeparttable}
\end{table*}

\newpage
\pagebreak [4]
$$\\$$
$$\\$$$$\\$$

\section{Modelfit parameters}\label{modelfit}
In this Section we report the modelfit parameters of the three jetted sources \txsB{}, \pks{} and \nvssB{}. These results are presented in  Sect.~\ref{result} and discussed in Sect.~\ref{discussion}. 

\begin{table*}[htb]
\centering
\begin{threeparttable}
\caption{Modelfit component parameters of \textbf{\txsB{}}.}
	\label{tab:mf_txs}
 \begin{tabular}{ c c c c c c c c c c c c}
 \hline \hline
Date & Obs. & $\nu$ & Comp. & Flux  &  Radius &  P.A.  & Maj. Axis & Ax. ratio &  10\%\,beam & $   \theta_{\mathrm{beam}}$\\ 
&& $\mathrm{(GHz)}$ &  &(mJy)   & (mas)   & (deg) &  (mas)  &      &\small $\mathrm{(mas \times mas)}$ &  (deg) \\
\small(1) & \small(2) &\small (3) &\small (4) &\small (5) &\small (6) & \small(7) &\small (8) &\small (9) &\small (10)  &\small (11)\\
 \hline 
  2004-04-30  & RFC & 2.3 GHz & 1  &  280$\pm$28 & 0.05 & 21 & 1.4 & 0.4 &   0.4$\times$0.3 & $-$0.1   \\
&& & 2 &2.4$\pm$0.2 & 4.0  &  128 & 4.7  &    0.7 & \\ 
&& & 3 & 10$\pm$1   & 10.0 & 149 &  2.4  &    0.5 & \\
&& & 4 & 1.8$\pm$0.2  & 15.2 & 141 & 11.0 & 0.5 &  \\
&& & 5  & 12$\pm$1 & 28.5 & 157 & 8.9  & 0.5 &   \\
2004-04-30  & RFC & 8.6 GHz & 1  & 287$\pm$29 & 0.1 & 105 & 0.4 & 0.6 &  0.1$\times$0.2 & 1.1  \\
&& & 2 &21$\pm$2 & 0.7 & 157 &  0.8 & 0.5$^{*}$ &  \\
&& & 3 &9.3$\pm$0.9 & 2.7 & 145  & 2.9  & 1.0$^{*}$  & \\
 2007-08-01 & RFC & 8.4 GHz & 1 & 384$\pm$38  & 0.006  &  62   &  0.5    & 1.0$^{*}$  &    0.2$\times$0.1 & 35.1  \\
&& & 2 &  22$\pm$2  & 1.2    & 159  &  1.0     &  1.0$^{*}$ &    \\
 2012-02-20 & RFC & 8.4 GHz & 1 & 100$\pm$10  & 0.09 &  $-28$   &  0.3    & 0.5  &  0.3$\times$0.1 & 8.0\\
&& & 2 &  4.7$\pm$0.5  & 1.0    & 148  &  0.5     &  0.98 &    \\
&& & 3 & 0.57$\pm$0.06  & 4.6    & 150 &  5.0    &  0.3 &    \\
           \hline
2020-02-29 & EVN & 4.9 GHz & 1  & 306$\pm$31   &  0.05  &   166  &   0.2   &  1.0$^{*}$  & 0.4$\times$0.3 & 7.8   \\
&& & 2 & 10.8$\pm$1.1   & 3.0 &   142  &  0.5    &  1.0$^{*}$   \\
&& & 3 &  2.9$\pm$0.3    &   7.0   &    152 &  0.5      & 1.0$^{*}$   \\ 
&& & 4 &  4.8$\pm$0.5  &   10.7   &   151 &  3.5    &  1.0$^{*}$    \\
&& & 5 &  4.8$\pm$ 0.5   &  29.4   &    158 &   4.4   &  1.0$^{*}$  \\  
&& & 6 &  0.23$\pm$0.02  &  32.7  &    165 &   9.1   &  1.0$^{*}$  \\  
2020-02-04 & VLBA & 8.4 GHz & 1  & 318$\pm$32 &  0.07 & $-$19   &   0.15    & 0.5$^{*}$ & 0.2$\times$0.1 &$-$6.3 \\ 
&& & 2 & 73.8$\pm$7.4  &  0.45 &  158  &  0.2    & 0.9   \\
&& & 3 & 7.6$\pm$0.8 &  1.6  &  156  &  0.6   & 0.5 \\ 
&& & 4 & 5.9$\pm$0.6  &  3.1   &  143  &   1.1   & 0.2  \\  
&& & 5 & 1.2$\pm$0.1  &  7.0   &  147  &   1.6  &  1.0$^{*}$   \\ 
&& & 6 & 3.9$\pm$0.4 &  10.2  &  157 &   5.2   &  1.0$^{*}$ \\  
&& & 7 & 1.8$\pm$0.2 &  18.8 &  170  &   2.9  &  1.0$^{*}$ \\ 
2020-02-04 & VLBA & 23.5 GHz & 1  & 338$\pm$34 & 0.01  & $-$22  & 0.15 & 0.5$^{*}$  & 0.08$\times$0.03 &$-$9.6 \\
&& & 2 &  50$\pm$5 &  0.2 &  159  & 0.2$^{*}$  & 0.9$^{*}$ \\
&& & 3 &  3.0$\pm$0.3 &  0.9 & 153  & 0.6$^{*}$  & 0.5$^{*}$   \\
&& & 4 &  1.8$\pm$0.2  & 2.8  & 142  & 1.1$^{*}$ & 0.2$^{*}$  \\
           \hline

 \end{tabular}
\begin{tablenotes}[para,flushleft]
\item  \textit{Notes:} Col. 1 - Date of observation; Col. 2 -  Origin of the observation: RFC/MOJAVE or our VLBA/EVN/e-MERLIN observations; Col. 3 -  Observation Frequency in GHz; Col. 4 - Components numbering. We assign  numbers to components as a guide to visualize them. However, same numbers at different epochs and frequencies are not referred to same components in the jet because the components  are not  identified  at all  epochs and frequencies.  Then, in the case in which   this numbering is taken as reference, it  must been interpreted to each dataset independently.
 Col. 5 -  Flux density in mJy; 
 Polar coordinates: Col. 6 -    radius in mas and Col. 7 -    P.A.  in degrees, measured from North through East  of the component's center with respect to the image central pixel;
Col. 8 -  FWHM of the component's major axis in mas;
Col. 9 -  axial ratio between FWHM major and minor axis of the component; Col. 10 -   10\% of the  image restoring beam (major and minor axis in mas) and Col. 11  -  restoring beam orientation (from North through East, in degrees) indicated as  reference for the components position uncertainty. Parameters marked with the $^{*}$ symbol are fixed during the fitting procedure.
\end{tablenotes}
\end{threeparttable}
\end{table*}


\begin{table*}[htb]
\centering
\begin{threeparttable}	
\caption{Model-fit component parameters of \textbf{\pks{}}.}
\label{tab:mf_pks}
 \begin{tabular}{ c c c c c c c c c c c c}
 \hline \hline
Date & Obs. & $\nu$ & Comp. & Flux  &  Radius &  P.A.  & Maj. Axis & Ax. ratio &  10\%\,beam & $   \theta_{\mathrm{beam}}$\\ 
&& $\mathrm{(GHz)}$ &        & (mJy)   & (mas)   & (deg) &  (mas)  &      &\small $\mathrm{(mas \times mas)}$ &  (deg) \\

\small(1) & \small(2) &\small (3) &\small (4) &\small (5) &\small (6) & \small(7) &\small (8) &\small (9) &\small (10)  &\small (11)\\
 \hline 
 2018-10-06 & MOJAVE & 15.3 & 1 & 485$\pm$25 &  0.04 &    143 & - &    -   &   0.2$\times$0.06 & 6.7 \\
&& & 2 &  128$\pm$7 &  0.2 &  $-$38 &  - &   -  \\
&& & 3 &  7.4$\pm$0.4 &  1.1 &  $-$27 &  0.4 &   1.0$^{*}$ \\
&& & 4 &  0.70$\pm$0.04 &  3.3 &  $-$26 &  0.17 &   1.0$^{*}$ \\
2019-07-19 & MOJAVE & 15.3 & 1 & 491$\pm$25 &  0.05 & $-$18 & 0.04 &   1.0$^{*}$   &   0.2$\times$0.06 & $-$18.1 \\
&& & 2 &   17$\pm$1 &  1.0 &  $-$33 &  0.3 &  1.0$^{*}$ \\
&& & 3 &   1.18$\pm$0.05 & 2.7  &  $-$24 & 0.9 &   1.0$^{*}$ \\
 2020-05-25& MOJAVE & 15.3 & 1 &   455$\pm$46 &  0.006 &  $-$32 &  0.17 &    0.09  &   0.2$\times$0.07 & $-$21.6 \\
&& & 2 &   13.1$\pm$0.1 &  0.7 &  $-$31 &  0.15 &  1.0$^{*}$ \\
&& & 3 &    2.6$\pm$0.2 &  2.8  &  $-$31 & 3.9  &   0.10 \\
2020-10-21 & MOJAVE & 15.3 & 1 &  532$\pm$53 &   0.01 &  $-$2.9 &  0.17 &   0.4  &   0.11$\times$0.05 & $-$4.6 \\
&& & 2 &   20$\pm$2 &   0.5  &   $-$26 &   0.5 &  0.4 \\
&& & 3 &   2.9$\pm$0.3 &  1.4  &   $-$31 &  0.5 & 1.0$^{*}$ \\
&& & 4 &   2.1$\pm$0.2 &  4.2  &   $-$24 &  1.5 &   1.0$^{*}$ \\
2020-12-01  & MOJAVE & 15.3 & 1 &   635$\pm$64   & 0.004 & 20  &  0.14  &  0.2   &   0.12$\times$0.06 & $-$4.8 \\
&& & 2 &   13.5$\pm$1.4   &  0.5 & $-$29 &   0.2  &  0.3 \\
&& & 3 &   4.73$\pm$0.5 & 1.3 &   $-$32 &  0.7  & 0.3 \\
&& & 4 &   0.9$\pm$0.09 & 2.6 & $-$30 &  0.7 &   1.0$^{*}$ \\
&& & 5 &   1.3$\pm$0.1 & 4.5  & $-$24 & 1.4 &  1.0$^{*}$ \\
         \hline
 2020-11-05 &  e-MERLIN & 5.1 & 1 & 333$\pm$17 &  0.2 &     $-$176 &  4.3 &    0.1   &  8.2$\times$3.9 & 23.9 \\
&& & 2 &  6.7$\pm$0.4&  285 &  $-$15 &  291 &    0.4   \\
&& & 3 &  2.8$\pm$0.2 &  703 &  $-$29 &  186 &   1.0$^{*}$ \\
         \hline
 \end{tabular}
\begin{tablenotes}[para,flushleft]
\item  \textit{Notes:} Same as Table~\ref{tab:mf_txs}.
\end{tablenotes}
\end{threeparttable}

\end{table*}


\begin{table*}[htb]
\centering
\begin{threeparttable}
\caption{Model-fit component parameters of \textbf{\nvssB{}}.} \label{tab:mf_0658} 
\begin{tabular}{ c c c c c c c c c c c c}
 \hline \hline
Date & Obs. & $\nu$ & Comp. & Flux  &  Radius &  P.A.  & Maj. Axis & Ax. ratio &  10\%\,beam & $   \theta_{\mathrm{beam}}$\\ 
&& $\mathrm{(GHz)}$ &        & (mJy)   & (mas)   & (deg) &  (mas)  &      &\small $\mathrm{(mas \times mas)}$ &  (deg) \\

\small(1) & \small(2) &\small (3) &\small (4) &\small (5) &\small (6) & \small(7) &\small (8) &\small (9) &\small (10)  &\small (11)\\
 \hline 
 2013-04-08/09 & RFC & 4.3 & 1 & 18.7$\pm$1.2 & 0.17 & 82  & 0.4 & 1.0$^{*}$ &  0.5$\times$0.2 &$-$7.3     \\
&& & 2 &  6.2$\pm$0.6 & 1.1 & $-$102  & 1.0 & 1.0$^{*}$  \\
 2013-04-08/09 & RFC &7.6   & 1  & 22.3$\pm$2.2 & 0.03 &     $-$59 &    0.4  &  1.0$^{*}$  &  0.3$\times$0.11 & $-$12.2  \\
&& & 2 &  3.5$\pm$0.4 &   1.6 &  $-$116    &     1.0 &    1.0$^{*}$  \\
2013-10-19 & RFC &  7.6 & 1 & 11.9$\pm$1.2 & 0.01  & 145  & 0.5 & 1.0$^{*}$ &   0.2$\times$0.13 &$-$3.3  \\
&& & 2 & 3.0$\pm$0.3 & 1.4 & $-$114 & 0.9 & 1.0$^{*}$ \\
&& & 3 & 1.1$\pm$0.1 & 2.7  & $-$119 & 1.4  & 1.0$^{*}$ \\
         \hline
2020-12-01/02 & EVN & 4.9 & 1 & 7.3$\pm$0.7 &  0.7 &     146 &    -  &    -   &    0.2$\times$0.11 & 82.4 \\
               &    &     & 2 & 5.5$\pm$0.6 &  1.2 &  $-$115 &  1.3 &    1.0$^{*}$ & &  \\
                 &    &     & 3 &    0.5$\pm$0.1  &  5.1 &  $-$79 &  6.9 &   1.0$^{*}$  & &  \\
 2020-12-06  & VLBA & 8.4 & 1 & 11$\pm$1 & 0.3 & 124  & 0.9 & 0.7 &   0.2$\times$0.1 & 2.2    \\
 & && 2 &  3.7$\pm$0.4 & 1.5 & $-$118  & 1.3  & 0.7  \\
 & & &3 &  0.17$\pm$0.02 & 2.7  & $-$94 & 0.9 & 1.0$^{*}$  \\
 2020-12-06  & VLBA & 23.5 &1 & 9.3$\pm$0.9 & 0.8  & 150  & 0.2 & 0.6 &   0.1$\times$0.04 &$-$13.5   \\
&& & 2 &  4.5$\pm$0.4 & 1.5$^{*}$ & $-$143  & 1.3  & 0.7$^{*}$  \\
         \hline
 \end{tabular}
\begin{tablenotes}[para,flushleft]
\item  \textit{Notes:} Same as Table~\ref{tab:mf_txs}.
\end{tablenotes}
\end{threeparttable}
\end{table*}

\end{document}